\documentclass[11pt]{article}
\def\ds{\displaystyle}

\usepackage{amsmath}
\usepackage{amsfonts}
\usepackage{amssymb}\usepackage{amsmath}
\usepackage{amsfonts}
\usepackage{amssymb}
\begin{document}
\pagestyle{headings}
\renewcommand{\thefootnote}{\alph{footnote}}

\title{Fields and Quantum Mechanics}
\author{Glenn Eric Johnson\\Oak Hill, VA.\footnote{Author is reached at: glenn.e.johnson@gmail.com.}}
\maketitle

{\bf Abstract:} The quantum field theories (QFT) constructed in [\ref{gej05},\ref{mp01}] include phenomenology of interest. The constructions approximate: scattering by $1/r$ and Yukawa potentials in non-relativistic approximations; and the first contributing order of the Feynman series for Compton scattering. To have a semi-norm, photon states are constrained to transverse polarizations and for Compton scattering, the constructed cross section deviates at large momentum exchanges from the cross section prediction of the Feynman rules. Discussion includes the incompatibility of canonical quantization with the constructed interacting fields, and the role of interpretations of quantum mechanics in realizing QFT.

\section{Introduction}

Quantum field theories (QFT) may be exotic mathematical constructs that resist discovery, or one of the conditions applied may be too strong to admit realizations of physical interest. Feynman series developments achieve phenomenological success although there is no demonstration that the developments are consistent with quantum mechanics [\ref{pct},\ref{bogo},\ref{wightman-hilbert}]. Lacking is a demonstration that QFT of interest have Hilbert space realizations.

The assumption here is that the lack of realizations for QFT of interest results from an overly constrained description for the physical problem. Motivated by particuliar constructions [\ref{gej05},\ref{mp01}] for nontrivial vacuum expectation values (VEV), this work continues study of explicit VEV included in a relaxation of established axioms for QFT, a relaxation concerning assertions for fields as Hilbert space operators. The constructed realizations for QFT result when the Wightman axioms [\ref{pct},\ref{bogo},\ref{wight},\ref{borchers}] are revised [\ref{gej05}] to permit additional sets of generalized functions. In the constructions, the algebra of functions labeling the states lacks the $*$-involution assumed in the Wightman axioms. Consequently, fields are not Hermitian Hilbert space operators. Acceptance of this contradiction to established assertions for QFT admits constructions of local, Poincar\'{e} covariant, positive energy states exhibiting interaction. These states are elements of Hilbert spaces.

To establish physical relevance and classical limits for the constructions, equivalent potentials and Feynman rules scattering amplitudes are contrasted with the constructed, explicitly quantum mechanical scattering amplitudes. The demonstrations are that:\begin{itemize}\item[--] the constructed QFTs include Yukawa and $1/r$ (as a limit of Yukawa) potentials as effective potentials in non-relativistic approximations.\item[--] a variation of the constructions approximates the Compton scattering cross section derived from the first contributing order of the Feynman series. The fractional error with respect to the Feynman series result becomes appreciable only when the product electron exhibits a relativistic velocity in the rest frame of the incident electron. To approximate Feynman rules results and achieve a semi-norm in the pre-Hilbert space, photons are restricted to transversely polarized states.\end{itemize}The constructed scattering amplitudes are scalar products of elements within Hilbert spaces and the demonstrable Hilbert space realization provides that scattering operators $U(t)^*U(-t)$ are unitary. Equality of cross sections allows scattering amplitudes to differ by a phase. Agreement of constructed and Feynman series scattering amplitudes up to a phase was demonstrated for weakly coupled ${:\!\Phi^4\!:}$ interaction in [\ref{gej05}]. The variations to the constructions are alternative compositions of submatrices from the two-point function matrix $M(p)$ into higher order connected functions.

The constructions provide examples of explicitly quantum mechanical developments that approximate established results. The examples demonstrate compatibility of the ``fundamental notions of the theory (relativistic invariance, quantum mechanics, local fields, etc.)'' [\ref{pct}], but the realizations are incompatible with canonical quantization. The achievement of interaction in the constructions precludes the $*$-involution of the algebra of function sequences. The established positivity axiom for QFT consists of two assertions:\begin{itemize}\item[--] the Wightman-functional provides a semi-norm for an algebra of function sequences;\item[--] the algebra has the $*$-involution.\end{itemize}The semi-norm results in the Hilbert space of states that realizes quantum mechanics. The $*$-involution is a technically convenient assumption, implying in particular that fields are Hermitian Hilbert space operators, but its physical motivation is not as compelling. Hermiticity of Hilbert space field operators can be disputed: in significant cases, position operators are Hermitian only in non-relativistic limits [\ref{wigner}]; quantities such as $x^3p$ are observable in classical limits of ordinary quantum mechanics but do not correspond to Hermitian operators [\ref{gej05},\ref{bogo}]; and Hermiticity is not decidable by observation. Abandonment of the $*$-involution provides example realizations for QFT exhibiting interaction. The history of QFT has been a demonstration of the difficulty, if not infeasibility, of realizing interaction in a QFT with fields that are self-adjoint operators in a Hilbert space of local, Poincar\'{e} covariant, positive energy states. It is questionable whether only selected quantities need correspond to self-adjoint operators and that those selected quantities must correspond to self-adjoint operators. The necessity of Hermitian Hilbert space field operators is eliminated in the revision to the QFT axioms. Hermitian Hilbert space field operators are allowed, and realized with free fields. The constructions are denoted ``quantum field theories'' even though the fields are not necessarily Hilbert space operators. Here, QFT refers to the synthesis of quantum mechanics and relativity rather than to ``quantization'' of classical field equations. The terminology of fields is maintained to introduce formal algebras of fields that underlie the constructions, and in anticipation that the predictions of QFT and classical field theories agree in appropriate limits. The constructed VEV satisfy the established QFT axioms except for the assertion that the algebra of function sequences used in construction of the Hilbert space has the $*$-involution.

Assertion that there are Hermitian field operators, or equivalently, that the underlying algebra of function sequences has a $*$-involution, is decisive to construction of QFT exhibiting interaction. The constructions provide local, positive energy, Poincar\'{e} covariant states in Hilbert spaces that exhibit particles in interaction but lack Hermitian field operators. The constructions are realizations of states appropriate for quantum mechanics. The issue of a $*$-involution is decisive since there are no known alternative Hilbert spaces of interest. At issue is whether observables must be self-adjoint operators and if not, what observable properties lack corresponding self-adjoint operators? The answer is clearly `not' as the $x^3p$ example in ordinary quantum mechanics and the position operator $x$ within a one particle subspace of relativistic quantum mechanics clarify. The issue then is what observables lack corresponding operators? When the field is not required to be a self-adjoint Hilbert space operator, then the constructions are viable as QFTs. Observation cannot eliminate the possibility that states only approximate eigenstates of operators associated with observables and that these eigenstates lie beyond limits of elements of the Hilbert spaces. Consequently, it is not appropriate to include such assertions in the axioms for QFT. It is this unsupported assertion that precludes realizations of QFT exhibiting interaction.

%= = = = = = = = = = fields and scalar product = = = = =
Fields $\Phi(x)_{\kappa}$ with $x\in {\bf R}^d$ and $\kappa\in \{1,2,\ldots N_c\}$ are elements of the ring of fields with the complex numbers. The QFT is defined by generalized functions that are formally VEV of the fields,\begin{equation}\label{sesquis2} \renewcommand{\arraystretch}{1.25} \begin{array}{rl} W_n((x)_n)_{\kappa_1 \ldots \kappa_n} &:= \langle \Omega| \Phi(x_1)_{\kappa_1}\ldots \Phi(x_n)_{\kappa_n} \Omega\rangle\\
 &\;= \langle \Phi(x_k)_{\kappa_k} \ldots \Phi(x_1)_{\kappa_1}\Omega| \Phi(x_{k+1})_{\kappa_{k+1}}\ldots \Phi(x_n)_{\kappa_n} \Omega\rangle \end{array}\end{equation}for each $k$. The VEV are components of the Wightman-functional [\ref{borchers},\ref{yngvason}],\[\underline{W}:=(1,W_{1,\kappa_1},W_2(x_1,x_2)_{\kappa_1,\kappa_2},\ldots),\]dual to terminating sequences of functions $\underline{f}:=(f_0,f_1(x_1)_{\kappa_1},f_2(x_1,x_2)_{\kappa_1,\kappa_2}\ldots) \in {\cal A}$. The sesquilinear function on ${\cal A}\times {\cal A}$,\[\renewcommand{\arraystretch}{2} \begin{array}{rl}\underline{W}(\underline{f}^* \,{\bf x}\, \underline{g}) &={\ds \sum_{n,m} \int} d(\xi)_{n+m}\;((D\cdot)_n \tilde{W}_{n+m}((\xi)_{n+m})) \overline{\tilde{f}_n((-\xi)_{n,1})}{\ds \prod_{k=n+1}^{n+m}} \tilde{g}_m((\xi)_{n+1,n+m})\\
 &= {\ds \sum_{n,m} \int} (d\xi)_{n+m}\; \langle \tilde{\Phi}_m\ldots \tilde{\Phi}_1 \Omega, \tilde{\Phi}_{m+1}\ldots \tilde{\Phi}_{m+n}\Omega\rangle \tilde{f}_m^*((\xi)_m) \tilde{g}_n((\xi)_{m+1,m+n}) \end{array}\]provides the scalar product of states. The notation is from Appendix \ref{app-vev}. $\underline{W}$ is positive semidefinite for the subalgebra ${\cal B} \subset {\cal A}$ consisting of functions that vanish on the negative energy support of the VEV. The bijective map of equivalence classes in ${\cal B}$ to a dense set of elements in the Hilbert space has the isometry,\begin{equation}\label{isometry}\langle \underline{f}|\underline{g}\rangle= \underline{W}(\underline{f}^* \,{\bf x}\, \underline{g})\end{equation}using\begin{equation}\label{prod} \underline{f} \,{\bf x}\, \underline{g} := (f_0g_0, \ldots, \sum_{\ell=0}^n f_{\ell}((x)_{\ell})_{\kappa_1\ldots \kappa_{\ell}} \,g_{n-\ell}((x)_{\ell+1,n})_{\kappa_{\ell+1}\ldots \kappa_n}, \dots)\end{equation}as the product in the algebra of function sequences ${\cal A}$. The $*$-mapped function sequences are defined using a Dirac conjugation matrix $D$.\begin{equation}\label{dualf}\tilde{f}_n((p)_n)^*_{\kappa_1\ldots \kappa_n} := \sum_{\ell_1,\ldots \ell_n} D_{\ell_1 \kappa_1}\ldots D_{\ell_n \kappa_n} \overline{\tilde{f}_n} (-p_n, -p_{n-1}, \ldots ,-p_1)_{\ell_n \ldots \ell_1}.\end{equation}The $*$-map (\ref{dualf}) is expressed compactly as\[((D^T \cdot)_n \overline{\tilde{f}_n} ((-\xi)_{n,1})):=\tilde{f}_n((p)_n)^*_{\kappa_1\ldots \kappa_n}.\]The $*$-map satisfies $\underline{f}^{**}=\underline{f}$, $(\lambda \underline{f})^* =\overline{\lambda} \underline{f}^*$, $(\underline{g}+\underline{f})^* =\underline{g}^* + \underline{f}^*$ and $(\underline{g} \,{\bf x}\, \underline{f})^* =\underline{f}^* \,{\bf x}\, \underline{g}^*$. The $*$-map (\ref{dualf}) is an involution when it is an automorphism. States are labeled by equivalence classes of function sequences in ${\cal B}$, equivalent in the semi-norm\begin{equation}\label{norm} \| \underline{f} \|_{\cal B}:= \sqrt{\underline{W}(\underline{f}^* \, {\bf x}\, \underline{f})}.\end{equation}With the support constraint for ${\cal B}$, the constructed Hilbert space includes only states of positive energy. These constructions satisfy the Wightman axioms except that ${\cal B}^* \neq {\cal B}$. The $*$-map (\ref{dualf}), an automorphism of ${\cal A}$, is not an automorphism of ${\cal B}$.

A free field is described by a two-point function determined from two complex-valued $N_c \times N_c$ matrices: $M(p)$ with elements that are multinomials in the energy-momentum components as a result of Lorentz covariance [\ref{bogo}], and the Dirac conjugation $D$. On ${\cal B}$,\begin{equation}\label{twopoint}\renewcommand{\arraystretch}{1.25} \begin{array}{rl} \tilde{W}_2(p_1,p_2)_{\kappa_1 \kappa_2} &:=\tilde{\Delta}(p_1,p_2)_{\kappa_1 \kappa_2}\\
 &:= \delta(p_1+p_2)\; \delta_2^+ \, M(p_2)_{\kappa_1 \kappa_2}\\
 &= \delta({\bf p}_1+{\bf p}_2)\; \delta_1^- \delta_2^+ \; 2\,\sqrt{\omega_1\omega_2}\; M(p_2)_{\kappa_1 \kappa_2}.\end{array}\end{equation}The matrices satisfy conditions\begin{equation}\label{matcond} \renewcommand{\arraystretch}{1.25} \begin{array}{rl}
DM(p) &= C^*(p) C(p)\\
\overline{D}D&=1\\
M(p) &= \left( \renewcommand{\arraystretch}{1} \begin{array}{cc} M_1(p) & 0\\ 0 & M_2(p)\end{array} \right)\end{array}\end{equation}and\begin{equation}\label{matlocal}M_k(-p)^T = (-1)^{k\!-\!1} M_k(p).\end{equation}(\ref{matcond}) implements a semi-norm for function sequences, and locality results from (\ref{matlocal}) with commutation applied in the $N_b \times N_b$ boson component $M_1$ and anticommutation applied in the fermion component $M_2$. (\ref{matcond}) implies that $M(p)^* =DM(p) D^T$. Poincar\'{e} covariance is implemented by\begin{equation}\label{matcond2} \renewcommand{\arraystretch}{1.25} \begin{array}{rl}
S(A)M(p)S(A)^T &= M(\Lambda^{-1} p)\\
\overline{S}(A) D&= DS(A)\end{array} \end{equation}with $S(A)$ an $N_c$-dimensional representation of the covering group of the proper orthochronous Lorentz group with $A\in$SL(2,C), and that the supports of the Fourier transforms of the VEV include only points with $p_1+p_2\ldots +p_n=0$. The representations $S(A)$ are generally reducible and then subspaces of states with the desired spin are selected. The VEV that exhibit interaction are constructed from $M(p)$, its submatrices, and Lorentz invariant functions. The notation and VEV from [\ref{gej05}] are described in Appendices \ref{app-vev} and \ref{app-trun}.

The relaxation of constraints on the VEV enables satisfaction of the physical Wightman conditions. The physical Wightman conditions are a suitable Hilbert space of positive energy states, Poincar\'{e} invariance, and microcausality. While the constructions share properties with canonical quantizations and derive from a revision of earlier axiomatic QFT developments, there are significant differences between the constructions and established developments:
\begin{enumerate}\item The development departs from conjecture that fields are unbounded Hilbert space operators. States are elements in a Hilbert space constructed from local, Poincar\'{e} covariant VEV. States are labeled by functions $\underline{f}\in {\cal B}$ with Fourier transforms that contribute only for positive energies. The VEV are generalized functions for an enveloping set of function sequences ${\cal A}={\cal A}^*$ with ${\cal B}\subset {\cal A}$. Restriction to ${\cal B}$ selects the physical states that satisfy spectral (positive energy) support conditions. The semi-norm applies in the smaller set of functions ${\cal B}$ and for the constructions provided, the semi-norm can be extended to ${\cal A}$ only when there is no interaction. The positive energy support limitation implies that the constructed states have unbounded spatial support, although support may be arbitrarily dominantly within a bounded region. The unbounded spatial support follows from [\ref{segal}] for massive particles, and in the case of massless particles, the unbounded spatial support is necessary to admit non-zero functions with Fourier transforms that vanish together with all derivatives at zero energy [\ref{mp01}]. ${\cal A}$ includes functions of bounded spacetime support and locality of the VEV is conventional within ${\cal A}$. Signed symmetry under transpositions of adjacent, space-like separated arguments results from two-point functions based on the Pauli-Jordan function [\ref{steinmann}] and higher-order connected functions that exhibit the signed symmetry regardless of the separation of arguments. Although the constructed fields are not self-adjoint Hilbert space operators, local symmetry of the VEV is necessary to eliminate correlation of fields at space-like displacements, with correlation extrapolated from VEV evaluated with bounded support elements of ${\cal A}$. Microcausality is a property of the VEV for bounded support elements in ${\cal A}$ that is extrapolated to ${\cal B}\subset {\cal A}$. As a result, states exhibit Bose-Einstein or Fermi-Dirac statistics. The restriction of ${\cal B}$ to positive energy contributions results in satisfaction of the spectral support condition and enables satisfaction of locality with symmetric generalized functions. The $*$-involution is not an involution for the algebra of positive energy function sequences. When $\underline{f}\in {\cal B}$, the $\underline{f}^*$ from (\ref{dualf}) is in the null space of the semi-norm (\ref{norm}) provided by the Wightman functional.
\item An illustrative set of VEV is provided by the inverse Fourier transforms of the connected functions for a single scalar field,\[{^C \tilde{W}}_n(p_1,\ldots p_n):= c_n \,\delta(p_1+\ldots p_n) \prod_{k=1}^n \delta(p_k^2-m^2).\]$n\geq 4$ for these ${^C \tilde{W}}_n$ that are supplemented with a free field two-point function and an energy ordering, (\ref{w-defn}) of Appendix \ref{app-trun}. The energy ordering provides that $E_k>0$ when $E_{k-1}>0$. These VEV are generalized functions when $d\geq 4$ [\ref{mp01}], and $d\geq 3$ suffices when all particles have finite mass [\ref{gej05}]. $d$ is the number of spacetime dimensions. Connected functions ${^C\!W}_n$ are identified as the connected contributions of the Wightman functions $W_n$. This elementary example of a realization for the VEV of a QFT is excluded by the Wightman axioms.
% = = = = = structure of A = = = = =
\item \label{foot-A} In the constructions, energies are on mass shells. As a result of this particular form, the Wightman-functional is essentially a functional of the momenta only, of one less dimension than the spacetime dimension $d$. The constructed VEV can be described from\begin{equation}\label{emsupp}\renewcommand{\arraystretch}{1.25} \begin{array}{rl} W(f) &= {\ds \int} dE\, d{\bf p}\; \delta(E\pm \omega) \tilde{T}({\bf p}) \tilde{f}_1(p)\\
 &={\ds \int} d{\bf p}\; \tilde{T}({\bf p}) \tilde{f}_1(\pm \omega, {\bf p})\end{array}\end{equation}for each of the multiple arguments and components. The topology in ${\cal A}$ derives from the topologies for the Fourier transform function spaces. The Fourier transform functions have $n$ $d$-dimensional energy-momentum arguments and are tempered functions of the momenta $({\bf p})_n \in{\bf R}^{n(d-1)}$ when evaluated on mass shells. $\tilde{f}_n((\pm\omega,{\bf p})_n)_{\kappa_1 \ldots \kappa_n} \in S({\bf R}^{n(d-1)})$. The Fourier transform functions have the topology of $S({\bf R}^{(d-1)n})$. $f_0 \in {\bf C}$. For a single argument,\[\tilde{f}_1(p):=\tilde{g}(p) \tilde{f}({\bf p})\]with $\tilde{f}({\bf p}) \in S({\bf R}^{d-1})$, a tempered test function, and $\tilde{g}(p)$ is a multiplier for $S({\bf R}^d)$ [\ref{gel2}]. $\tilde{g}(p)$ together with all derivatives vanishes at $E=0$ when $m_\kappa=0$. The elements of ${\cal B}\subset {\cal A}$ have\[\tilde{f}_n((-\omega,{\bf p})_n)_{\kappa_1 \ldots \kappa_n}=0\]if for any $k\in \{1,\ldots n\}$ and $n>0$. In (\ref{emsupp}), both $T({\bf x})$ and $\tilde{T}({\bf p})\in S'({\bf R}^{d-1})$, generalized functions dual to tempered functions. The component functions $f_n((x)_n)_{\kappa_1 \ldots \kappa_n}$ from sequences in ${\cal A}$ are $d$ dimensional inverse Fourier transforms, convolutions\[f_1(x)=\int \frac{d{\bf y}}{(2\pi)^{\frac{d\!-\!1}{2}}}\; g(t,{\bf y})f({\bf x}-{\bf y})\]of the Fourier transform as a generalized function of the multiplier $g(x) \in S'({\bf R}^d)$ and the tempered function $f({\bf x})\in S({\bf R}^{d-1})$. These spacetime realizations are equivalent to tempered functions for the tempered packet states, and are generalized functions with point support in time when $\tilde{g}(p)$ is a polynomial in $E$.
\item ${\cal B}$ includes only trivial sequences of  real functions. Real functions from a sequence in ${\cal B}$, for example, $\tilde{f}_1(p)=(p^2-m^2)f({\bf p})$, are in the equivalence class of zero for the semi-norm (\ref{norm}). Only $f_0$ can be real. The lack of contributing functions in ${\cal B}^* \cap {\cal B}$ precludes the demonstration that multiplication of function sequences defines fields as Hilbert space field operators, $\Phi(\underline{f})\underline{g}:= \underline{f}\,{\bf x}\,\underline{g}$ [\ref{borchers}].\[\langle \underline{h}|\Phi(\underline{f})\underline{g}\rangle:= \underline{W}(\underline{h}^* \,{\bf x}\, (\underline{f}\,{\bf x}\,\underline{g}))= \underline{W}((\underline{f}^*\,{\bf x}\,\underline{h})^* \,{\bf x}\, \underline{g}).\]Cauchy's inequality would provide that the field preserves equivalence classes but Cauchy's inequality does not follow for the indefinite form $\underline{W}((\underline{f}^*\,{\bf x}\,\underline{h})^* \,{\bf x}\, \underline{g})$. Generally, $\underline{f}^*\,{\bf x}\,\underline{h} \not \in {\cal B}$. When the semi-norm does not extend beyond ${\cal B}$, the constructed interacting fields are not Hermitian Hilbert space operators.
\item \label{foot-notatn}The $*$ notation appears in three contexts: Hilbert space operator adjoints $\Phi(\underline{f})^*$; $*$-mapped function sequences $\underline{f}^*$; and formal adjoints of free field operator-valued distributions $\Phi_o(x)^*$. When there is a Hilbert space field operator, then $\Phi(\underline{f})^* = \Phi(\underline{f}^*)$ results from $\Phi(\underline{f})\,\underline{g}:=\underline{f}\,{\bf x}\,\underline{g}$ with the $*$-map (\ref{dualf}) and scalar product (\ref{isometry}). Here the definition for field uses a single argument,\[\underline{f}=(0,f(x)_1,\ldots f(x)_{N_c},0,0,\ldots),\]and $\Phi(\underline{f})$ is the sum $\Phi(\underline{f})=\sum_\kappa \Phi(f_\kappa)_\kappa$. $\Phi(\underline{f})^*$ is the Hilbert space operator adjoint of $\Phi(\underline{f})$ and $\Phi(\underline{f}^*)$ is the Hilbert space field operator labeled by $\underline{f}^*$. Hermiticity is that\[\Phi(\underline{f})^* = \Phi(\underline{f}).\]In free field developments, $*$-mapped functions are often considered complex conjugates, $\underline{f}(x)\mapsto \overline{\underline{f}}(x)$, and the adjoints of operator-valued distributions considered as the formal $\Phi_o(x)\mapsto \Phi_o(x)^*$. Here, the $*$-mapped function sequences are $\underline{f}(x)\mapsto \underline{f}(x)^*:=D^T\, \overline{\underline{f}}(x)$ with fields that are formally Hermitian, $\Phi(x)\mapsto \Phi(x)$ from (\ref{sesquis2}). In the example of a charged field, $\Phi=(\Phi_a,\Phi_b)=(\Psi,\Psi^*)$, $f=(f_a,f_b)$ and $f^*=(\overline{f_b},\overline{f_a})$. In this case, Hermiticity is\[\Phi(\underline{f})^*=\Phi(\underline{f}^*)=\Psi(\overline{f_b})+\Psi(\overline{f_a})^*=\Phi(\underline{f})=\Psi(f_a)+\Psi(f_b)^*.\]Hermiticity is precluded in the constructions by a lack of real functions. $f_a= \overline{f_b}$ is not satisfied in ${\cal B}$. Formally $\Phi(x)^*_a=\Phi(x)_b=\Psi(x)^*$ but the lack of nontrivial real functions precludes Hermitian Hilbert space operators when there is interaction. $\Phi(\underline{f})^* = \Phi(\underline{f}^*)$ is implemented for operator-valued distributions by $\Phi(x)^*=D\Phi(x)$ when it is asserted that $\underline{f}^*=\overline{\underline{f}}$.
\item With the revised axioms and considered as boundary values of analytic functions, the VEV include terms with differing domains of holomorphy.
\item Lacking the $*$-involution, nontrivial interaction is consistent with a two-point function in the form of a free field two-point function, a Pauli-Jordan function and its generalizations to greater spin. The constructed local fields are not Hermitian Hilbert space operators and therefore are not symmetric operators. The Jost-Schroer and similar theorems [\ref{bogo},\ref{feder},\ref{greenberg}] do not apply since the theorems' assumptions for Hilbert space field operators are violated by the constructions that exhibit interaction and violate the assertion of a $*$-involution from the Wightman axioms,.
\item The constructions exhibit interaction. $\langle U(t)\underline{f} | U(-t)\underline{g}\rangle \neq 0$ as $t\rightarrow \infty$ with plane-wave `in' states $|U(-t)\, \underline{g}\rangle$, `out' states $|U(t) \underline{f}\rangle$ and unitary time translation $U(t)$ unless $\langle \underline{f} | \underline{g}\rangle=0$.
\item The revised axioms are weaker than assumptions for local observables in the Araki-Haag-Kastler algebraic development of QFT. No assumptions are included in the revised axioms concerning association of subalgebras of Hilbert space operators with bounded subsets of spacetime. Projections onto subspaces of states and the generators of the Poincar\'{e} group are self-adjoint operators necessarily included in the constructions, but these operators are not associated with bounded subsets of spacetime. States labeled by functions from ${\cal B}$ include states localized in time, but not in space. The causal complement of the support of any function from $\underline{f}\in {\cal B}$ is empty. The free field semi-norm extends from ${\cal B}$ to include ${\cal A}$ with functions of bounded support but the extension does not apply when interaction is exhibited. Demonstrations that QFT with local observables, including free QFT, are type III Hilbert spaces does not evidently apply to the constructions of interest and the structure of the constructed Hilbert spaces has not been resolved. The lack of functions of bounded support precludes probabilities that vanish within extended regions, to much the same effect as type III factors [\ref{type3}].\end{enumerate}The constructions depart from established developments of QFT but are anticipated to be conventional as quantum mechanics: isolated states are labeled by attributes associated with free particles; the likelihood that a system is observed in a state $s$ is Trace($\rho P_s$) for a state density matrix $\rho$ and orthogonal projection operator $P_s$; states evolve as $U(t)\rho \,U(t)^*$ for a unitary $U(t)$; and the squared magnitudes of the scalar product of elements from the Hilbert space, for example $|\langle s'|U(t) s\rangle|^2$, are transition probabilities.

Appendices summarize the constructions [\ref{gej05},\ref{mp01}] and provide a background on electrodynamics suited to the constructions.

\section{States, observables and self-adjoint operators}

The constructions motivate consideration of alternatives to established developments for QFT and anticipate richer dynamical descriptions than ``quantization'' of laws derived from classical concepts. The constructed interacting fields necessarily lack properties required for a canonical quantization.

The Feynman rules to evaluate scattering amplitudes for QFT result from a canonical quantization,\[ \mbox{Classical Dynamics} \stackrel{\scriptstyle \mathit{``Quantization"}}{\Longrightarrow} \mbox{realization of quantum mechanics},\]that is in analogy with non-relativistic quantum mechanics. One development is to unitarily relate fields using distinct Hamiltonians for free and interacting fields. A correspondence of QFT with classical field theory provides the Hamiltonian forms but the Haag and Hall-Wightman-Greenberg theorems [\ref{pct},\ref{bogo}] indicate that the definitions for Hilbert space operators are inconsistent with interaction. In contradiction to a canonical quantization, the constructed generators of time translation are forms canonically associated with free fields. The constructions exhibit nontrivial dynamics with ``trivial'' Hamiltonians (\ref{constr-hamil}). This result would be precluded by the Jost-Schroer and similar theorems if fields were Hermitian Hilbert space operators. The loss of involution is decisive to admitting the constructions. From the perspective of the constructions, the Jost-Schroer theorem provides that only the physically trivial free fields have Hermitian Hilbert space field operators.

It is well established that canonical quantization does not naturally extend to field theories. The Stone-von Neumann theorem [\ref{svn}] does not apply for an infinite number of degrees of freedom, the Haag and Hall-Wightman-Greenberg theorems [\ref{pct},\ref{bogo}] apply, and deficiencies in the correspondence of classical quantities with operators [\ref{vonN}] persist.

This study employs only general principles of quantum theory: a complete set of positive energy states, relativistic invariance, and microcausality. Classical limits are left to be discovered. The constructions demonstrate alternatives within quantum mechanics to ``quantization'' of classical idealizations. With explicit constructions for VEV that exhibit interaction, this development is\[\mbox{Hilbert space realization of quantum mechanics}\stackrel{\scriptstyle \mathit{Limits}}{\Longrightarrow}\mbox{Classical Dynamics}.\]Quantum mechanics lacks the concepts of identifiable objects and trajectories except when classical limits apply. And, flaws in classical limits, for example, descriptions of the radiation reaction force in classical electrodynamics [\ref{jackson}], should be corrected in the quantum development. The early development of QED included anticipation that a formulation based upon the development of Maxwell and Lorentz would be flawed due to such difficulties with the classical theory [\ref{pct}]. That the very small and very large scale properties of states are determined by classical limits is suspect. Additional structure, structure without classical analogue, may not be exhibited in the classical limits. Insight from the continuous evolution of mean values, analogues of the association of Newton's equation with quantum mechanics using Ehrenfest's theorem [\ref{messiah}], applies when particle production is precluded.

The technical revision for QFT relates naturally with an interpretation of measurement as the determination of state relative to an observer [\ref{ewg}] and of observables as descriptions of those states. Here, observables are generally dissociated from self-adjoint Hilbert space operators. This dissociation enables significant simplifications to the mathematical description of QFT while the necessary physical content is preserved. In this interpretation, states associated with the description ``a particle located at $x$'' are not necessarily eigenstates of a self-adjoint Hilbert space location operator, nor even states with support limited to a bounded region of spacetime. Descriptions need only be appropriate for the states. A location $x$ is appropriately associated with a state when the state is predominantly supported near $x$. The association is state, an element of the Hilbert space, with location, the classical configuration space concept.

States are elements of Hilbert spaces\footnote{The more general state description is a state density matrix (statistical operator [\ref{vonN}]), a nonnegative self-adjoint operator of trace class [\ref{gleason}] or generalizations. The self-adjoint state density matrix decomposes as a linear combination of projections onto states.} and their descriptions are interpretations of measurements, interpretations that include the classical concepts of location as a point and field strength at a point. Interactions that correlate (entangle) states of an observer with states of interest are observations. Projections [\ref{birk}] onto subspaces of states and the generators of the Poincar\'{e} group are {\em observables}\footnote{From [\ref{dirac}], ``...call a real dynamical variable whose eigenstates form a complete set an {\em observable}.'' The Hilbert space operator terms used here are Hermitian, symmetric, and self-adjoint: an operator $A$ with domain ${\cal D}_A$ in a Hilbert space with scalar product $\langle u,v \rangle$ is {\em Hermitian} if $\langle u,Av\rangle= \langle Au,v \rangle$ for every $u,v \in {\cal D}_A$; a Hermitian operator is {\em symmetric} if ${\cal D}_A$ is dense; and a symmetric operator is {\em self-adjoint} if ${\cal D}_A={\cal D}_{A^*}$ and $Au=A^*u$ for every $u\in {\cal D}_A$. Then Hermiticity is necessary to symmetry and self-adjointness. An appropriate definition for a {\em real dynamical variable} is substantial to a resolution to QFT difficulties.} necessarily included in the constructions. Lack of an associated self-adjoint operator does not exclude a quantity from physical relevance. Indeed, position is relevant in relativistic quantum mechanics. The spectral (abstract kernel) theorem for rigged Hilbert spaces (Gelfand triples) [\ref{gel4}] provides that there is a decomposition of self-adjoint operators as real linear combinations of projections. This justifies the definition of {\em observable} for self-adjoint Hilbert space operators. But more generally, quantities in the descriptions of states, particularly the classical idealizations of position and field strength, do not necessarily correspond to self-adjoint operators. In the revised QFT axioms, no constraints are applied to realize particular self-adjoint Hilbert space operators.
% \footnote{And finite linear combinations using complex constants on joint domains. The range of convergent sequences of observables is well beyond the scope of this note.}
% The Hilbert space projection theorem provides a decomposition of Hilbert spaces into subspaces and orthogonal complements.

Although Feynman series have provided a conceptual framework for QFT and described, for example, the Lamb shift, a direct interpretation for the series as quantum mechanics appears intractable. The approach here is to avoid the constraints that follow from the assertion that fields are Hermitian Hilbert space operators. The assertion of Hermitian field operators extrapolates successful descriptions for free or non-relativistic objects to objects that are both relativistic and exhibit interaction. But this extrapolation is in spite of significant differences between ordinary and relativistic quantum mechanics and that, even in ordinary quantum mechanics, a correspondence of classical observables with self-adjoint operators is not the general case. Here, an appropriate, positive energy Hilbert space realization, Poincar\'{e} covariance and locality are considered more strongly motivated conditions than the $*$-involution and consequent Hermitian Hilbert space field operators.

Isolated systems, those with negligible spatial overlap with other objects, are described by a classical particle limit\footnote{The concern here is the approximation of quantum mechanics by classical mechanics and not the $\hbar \rightarrow 0$ limit of quantum mechanics.} as long as the state remains isolated.\footnote{Overlap of the {\em isolated} system with remote systems is neglected. And, any entanglement of an isolated system with remote states due to prior interactions is not captured in the classical description.} The constructions share this limit with established developments of QFT. A classical field limit for the constructions is obscured by the lack of $*$-involution due both to the inapplicability of canonical quantization and the loss of a semi-norm on real functions except in the case of free fields. Generalized random processes, associated with the equal-time fields when the algebra of function sequences has the $*$-involution, are definite forms for real functions. An indirect association with classical fields results from the approximation of results with strong phenomenological justification: the constructions approximate Feynman rules scattering amplitudes; and the constructions approximate scattering amplitudes from non-relativistic, potential scattering.

\section{Scattering amplitudes}

%= = = = = = = = = = LSZ states = = = = =
Scattering amplitudes are large time difference, plane wave limits of state transition amplitudes. In [\ref{gej05}], the scattering amplitudes are developed for the plane wave limits of LSZ (Lehmann-Symanzik-Zimmermann) states labeled by functions based on\begin{equation}\label{LSZ}\tilde{\ell}(t;p_j)_{\kappa_j}=e^{i\omega_jt}(\omega_j + E_j) \tilde{f}({\bf p}_j)_{\kappa_j}\end{equation}with $\tilde{f}({\bf p})_\kappa$ in $S({\bf R}^{d-1})$ [\ref{gel2}] and a parameter $t$. These functions are elements of ${\cal B}$ for finite mass, generalized functions with point support in time, and spatial test functions. For an LSZ state labeled by $\ell(t)_\kappa$,\begin{equation}\label{LSZ-spt}\renewcommand{\arraystretch}{1.75} \begin{array}{rl} U(-t)\Phi(\ell(t))U(-t)^{-1}&=i {\ds \int} d{\bf x}\; \hat{f}(t,{\bf x}) \stackrel{\leftrightarrow}{\partial}_o \Phi(t,{\bf x})\\
 &= {\ds \int} dp\; (\omega+E) e^{i(\omega-E)t}\tilde{f}({\bf p})\;\tilde{\Phi}(p)\\
 &={\ds \int} dp\; (\omega+E) \tilde{f}({\bf p})\;\tilde{\Phi}(p)\\
 &=\Phi(\ell(0))\end{array}\end{equation}with $t$ arbitrary and $\Phi(x)$ indicating a VEV argument using (\ref{sesquis2}) and not a Hilbert space field operator. The independence from $t$ results from the limitation of the support of the VEV to mass shells (\ref{emsupp}). $f \stackrel{\leftrightarrow}{\partial}_o g:= f\dot{g}-\dot{f}g$ and\[\hat{f}(x)_{\kappa} =\frac{1}{(2\pi)^{\frac{d\!-\!2}{2}}}\int d{\bf p}\; e^{i\omega t}e^{-i{\bf p}\cdot {\bf x}}\tilde{f}({\bf p})_{\kappa}\]is a smooth solution of the Klein-Gordon equation.

Evaluation of the plane wave limit scattering amplitudes below uses\[\tilde{f}({\bf p})_{\kappa} =\delta_L({\bf p}-{\bf q})\,w(p)_\kappa\]with\begin{equation}\label{LSZdelta} \delta_L ({\bf p} -{\bf q})= \left(\frac{L}{\sqrt{\pi}} \right)^{d-1} \; e^{-L^2 ({\bf p} -{\bf q})^2}>0\end{equation}and $w(p)_\kappa$ a multiplier [\ref{gel2}]. Support is concentrated near the momentum ${\bf q}$ in the plane wave limit as $L \rightarrow \infty$. $w(p)_\kappa$ specifies particle species and polarization independently of the plane wave limit. Plane wave ``in'' states are\begin{equation}\label{instates} \lim_{\stackrel{L \rightarrow \infty}{t\rightarrow -\infty}} |U(-t) \tilde{\ell}_n(t) \rangle \rightarrow |(q,w)_n^{\mathit{in}} \rangle \end{equation}and ``out'' states are the $t\rightarrow \infty$ limits.

The designation of two-in, two-out amplitudes follows the convention,\begin{equation}\label{M-defn}\langle (p,w)_{1,2}^{\mathit{out}} |(p,w)_{3,4}^{\mathit{in}} \rangle :=2\pi i \, \delta(p_1+p_2-p_3-p_4) {\cal M}(({\bf p},w)_4)\end{equation}with ${\cal M}$ a generalized function of $({\bf p},w)_4$ with each energy on a mass shell, $E_j=\omega_j$. ${\cal M}$ corresponds to the renormalized result from the Feynman rules for evaluation of scattering amplitudes [\ref{weinberg}] when the normalization for polarization descriptions $w_k$ is\begin{equation}\label{pold}\overline{w}_k^T DM(p_k) w_k=\frac{1}{2E_k}\end{equation}for each $k$ and with the energy on the mass shell, $E_k=\omega_k$. The equivalence with box normalization is in Appendix \ref{app-trans}.

For the VEV from [\ref{gej05}] and LSZ states, the evaluation of transition amplitudes for the two-in, two-out, non-forward amplitudes in the plane wave limit results in\begin{equation}\label{scatt-eval} \renewcommand{\arraystretch}{1.25} \begin{array}{l}{\cal M} =-i(2\pi)^{d-1} c_4 \;{\ds \frac{|\varsigma_2|^2}{4} \sum_{\kappa_1\ldots \kappa_4}} \, \overline{w_1}(p_1)_{\kappa_1}\overline{w_2}(p_2)_{\kappa_2}w_3(p_3)_{\kappa_3}w_4(p_4)_{\kappa_4} {\bf S}_{1,2}[{\bf S}_{3,4}[ \times\\
 \quad \overline{U_2(p_1\!-\!p_2) M(p_1\!-\!p_2)_{\kappa_1 \kappa_2}}\,U_2(p_3\!-\!p_4) M(p_3\!-\!p_4)_{\kappa_3 \kappa_4}\\
 \quad + \beta_2\Upsilon(p_1\!+\!p_3) (DB)(p_1\!+\!p_3)_{\kappa_1 \kappa_3}\beta_4\Upsilon(p_2\!+\!p_4) (DB)(p_2\!+\!p_4)_{\kappa_2 \kappa_4}\\
 \quad + \beta_3\Upsilon(p_1\!+\!p_4) (DB)(p_1\!+\!p_4)_{\kappa_1 \kappa_4}\beta_3\Upsilon(p_2\!+\!p_3) (DB)(p_2\!+\!p_3)_{\kappa_2 \kappa_3} ]].\end{array}\end{equation}The VEV are provided in (\ref{trun-eval}) of Appendix \ref{app-vev}. In the scattering amplitude (\ref{scatt-eval}), energies are on mass shells, and polarizations and species are described by $w_j(p_j)_{\kappa_j}$. In the plane wave limit (\ref{LSZdelta}), the only contributing term to the non-forward transition amplitudes is the connected function ${^C\!W}_n$. Corrections of order $L^{-1}$ that result from approximating the relatively slowly varying functions as constant within momentum summations are neglected. This approximation follows from the mean value theorem for integration given the nonnegativity of the delta sequences used in the LSZ functions (\ref{LSZdelta}), continuity of $M(p)$, $B(p)$, $\Upsilon(p)$, $U_k(p)$, $\omega_j$, and a regularization of the energy conserving delta, $\delta(E)$. The regularization is for convenience and limits the range of the summation over time to $-T/2$ through $T/2$.\[\int_{-T/2}^{T/2} du_{0}\; e^{-iEu_{0}}=\frac{2\sin(ET/2)}{E}\approx 2\pi \delta(E).\]The regularization is a convenience and not necessary to the result (\ref{scatt-eval}). This regularization of the energy-conserving delta and the summation over the LSZ functions (\ref{LSZdelta}) results in replacement of the energy-momentum conserving delta of (\ref{M-defn}) by the delta sequence\begin{equation}\label{delta} \renewcommand{\arraystretch}{1.75} \begin{array}{rl}
\delta_T(p;L^2/n) &:= {\ds \frac{\sin(ET/2)}{\pi E}}\,{\ds \int \frac{du}{(2\pi)^{d\!-\!1}} \int} d({\bf q})_n\; \left({\ds \frac{L}{\sqrt{\pi}}}\right)^{n(d-1)} {\ds \prod_{j=1}^n}\, e^{-L^2({\bf q}_j-{\bf p}_j)^2 -i{\bf q}_js_ju} \\
 &= {\ds \frac{\sin(ET/2)}{\pi E}} \left({\ds \frac{L}{\sqrt{n\pi}}}\right)^{d-1} e^{-L^2{\bf p}^2/n}\end{array} \end{equation}with $n=4$ for two-in, two-out processes. $p=\sum s_j p_j$ with $s_j=1$ for incoming and $s_j=-1$ for outgoing particles.

% = = = = = = Equivalent potentials = = = = = =
\section{Equivalent potentials}

In selected cases, nonrelativistic limits of scattering cross sections derived from the scattering amplitudes (\ref{scatt-eval}) are described by equivalent potentials. An equivalent nonrelativistic potential is a concept best adapted to description of the interaction of two distinguishable particles, that is, to a description with a classical limit. For indistinguishable particles, the exchange effect [\ref{rgnewton}] is exhibited in quantum mechanics and has no classical analog. There is no classical limit of the indistinguishability of particles and an effective potential is established for cases with particles imagined as distinguishable. Distinguisable particles may have distinct masses or differ by quantum numbers not involved in the interaction described by the equivalent potential.

The method is to compare the constructed scattering amplitudes (\ref{scatt-eval}) with the scattering predictions from ordinary quantum mechanics when an incident plane wave scatters due to interaction with a second, distinguishable particle. The interaction is described by a potential energy. The scattering is elastic and the interaction is between a particle species and polarization described by $w_j(p_j)_{\kappa_j}$ with a distinct particle species or polarization described by $w_k(p_k)_{\kappa_k}$. In this development, incoming particles designated 1 and scattering product 3 are described with the same species and polarization, and incoming particle 2 and scattering product 4 are similarly described by the same species and polarization that describe distinguishable states with regard to 1. That is, $w_1(p_1)_{\kappa_1}=w_3(p_1)_{\kappa_1}$, $w_2(p_2)_{\kappa_2}=w_4(p_2)_{\kappa_2}$ and $w_1(p_1)_{\kappa_1} \neq w_2(p_1)_{\kappa_1}$. 

With these understandings, (\ref{scatt-eval}) reduces to the form of a scattering amplitude for a single scalar field but with distinguishable particles. Evaluation of the scattering amplitude (\ref{M-defn}) results in\begin{equation}\label{scalar-amp} {\cal M}={\ds \frac{-i(2\pi)^{d-1} c_4}{\sqrt{\omega_1\omega_2\omega_3\omega_4}}}\left[
\overline{U_s(p_1\!-\!p_2)}\,U_s(p_3\!-\!p_4)+\hat{\Upsilon}_a(p_1\!+\!p_3,p_2\!+\!p_4)+ \hat{\Upsilon}_b(p_1\!+\!p_4,p_2\!+\!p_3)) \right]\end{equation}with\[U_s(p_i\!-\!p_j) := \sqrt{\omega_i\omega_j}\,\sum_{\kappa_i, \kappa_j} \, w_i(p_i)_{\kappa_i} w_j(p_j)_{\kappa_j}{\bf S}_{i,j}[ U_2(p_i\!-\!p_j) M(p_i\!-\!p_j)_{\kappa_i \kappa_j}]\]and\[ \renewcommand{\arraystretch}{1.25} \begin{array}{l} \hat{\Upsilon}_a(p_1\!+\!p_3,p_2\!+\!p_4)+ \hat{\Upsilon}_b(p_1\!+\!p_4,p_2\!+\!p_3)):= {\ds \sum_{\kappa_1\ldots \kappa_4}} \, \overline{w_1}(p_1)_{\kappa_1}\overline{w_2}(p_2)_{\kappa_2}w_3(p_3)_{\kappa_3}w_4(p_4)_{\kappa_4}\times\\
 \qquad \sqrt{\omega_1\omega_2\omega_3\omega_4}\,{\bf S}_{1,2}[{\bf S}_{3,4}[ \beta_2\Upsilon(p_1\!+\!p_3) (DB)(p_1\!+\!p_3)_{\kappa_1 \kappa_3}\beta_4\Upsilon(p_2\!+\!p_4) (DB)(p_2\!+\!p_4)_{\kappa_2 \kappa_4}\\
 \qquad \qquad \qquad \qquad +\beta_3\Upsilon(p_1\!+\!p_4) (DB)(p_1\!+\!p_4)_{\kappa_1 \kappa_4}\beta_3\Upsilon(p_2\!+\!p_3) (DB)(p_2\!+\!p_3)_{\kappa_2 \kappa_3} ]].\end{array}\]$|\varsigma_2|^2=1$. When $U_s(p)$, $\hat{\Upsilon}_a(p)$ and $\hat{\Upsilon}_b(p)$ are constants, this amplitude coincides with the first contributing order of weakly coupled $:\! \Phi^4\!:$.

The non-relativistic limit of the cross section, (\ref{elascs}) from Appendix \ref{app-trans}, is contrasted with the cross section from ordinary quantum mechanics. Using separation of variables in Jacobi coordinates, the center of mass motion is isolated from the relative motion of the two incoming particles, and evaluation of the cross section results from the scattering of a single particle of reduced mass reacting to a potential. The first Born approximation to the resulting Fredholm equation of the second kind, applicable when wave packet dispersion is negligible, at asymptotic distances and for potentials of rapid decline, results in [\ref{merzbacher}]\[\frac{\ds d\sigma}{\ds d\Omega} = \left|-\frac{\mu}{2\pi} \int d{\bf x}\; e^{i({\bf p}_1-{\bf p}_3)\cdot {\bf x}}\;V({\bf x})\right|^2\]for three spatial dimensions, $d=4$. ${\bf p}_1$ is a selected incident momentum and ${\bf p}_3,{\bf p}_4$ are the scattered momenta with ${\bf p}_3$ the scattered momentum of the particle of the same description as ${\bf p}_1$ in this case of distinguisable particles. \[\mu = \frac{m_1m_2}{m_1+m_2}\]is the reduced mass for particle masses $m_1,m_2$. For a mollified $1/r$ potential, this first Born approximation cross section coincides up to a phase with the ordinary quantum mechanical and classical Rutherford cross section evaluations [\ref{rgnewton}].

% = = = = = = = = = = = = = = = = = = = = = = = = = = = = = = = = = = = 
Narrowing consideration to the center of momentum frame and equal masses, $m_2=m_1$, $\mu=m_1/2$, the incoming momenta are $p_1 = (\omega_1,{\bf p}_1)$ and $p_2 = (\omega_1,-{\bf p}_1)$, and the outgoing momenta are $p_3 =(\omega_1,{\bf p}_1-{\bf q})$ and $p_4 =(\omega_1,-{\bf p}_1+{\bf q})$ with ${\bf q}:={\bf p}_1-{\bf p}_3$, a momentum transfer. The momentum dependence of the constructed scattering amplitude (\ref{scalar-amp}) follows from the identifications\[\renewcommand{\arraystretch}{1} \begin{array}{rlrl} p_3-p_4 &= (0,2{\bf p}_1-2{\bf q}),\qquad &p_1-p_2 &= (0,2{\bf p}_1)\\
p_3+p_1 &= (2\omega_1,2{\bf p}_1-{\bf q}),\qquad &p_4+p_2 &= (2\omega_1,-2{\bf p}_1+{\bf q})\\
p_3+p_2 &= (2\omega_1,-{\bf q}),\qquad &p_4+p_1 &= (2\omega_1,{\bf q}) .\end{array}\]In a non-relativistic limit, $\omega_k\approx m_1\gg \|{\bf p}_k\|$ and $(p_j+p_k)^2\approx 4m_1^2$. The contributions of the two terms in (\ref{scalar-amp}) including $\hat{\Upsilon}_\alpha(p)$ are nearly independent of the momenta in the nonrelativistic limit and their contribution to the equivalent potential is consequently relatively short range. The contributions of these two terms are neglected below.

Setting the two expressions for cross section equal results in\[\frac{\ds d\sigma}{\ds d\Omega} = \left|-\frac{\mu}{2\pi} \int d{\bf x}\; e^{i({\bf p}_1-{\bf p}_3)\cdot {\bf x}}\;V({\bf x})\right|^2= \frac{\ds (2\pi)^4 \varrho_o \, \omega_1\omega_2\omega_3\omega_4\, |{\cal M}(({\bf p},w)_4)|^2}{\ds \varrho_3\, (\omega_3+\omega_4)^2}.\]With equal masses, $\varrho_o=\varrho_3$ from (\ref{varrho}) of Appendix \ref{app-trans}. Then, in a non-relativistic limit, ${\bf p}_j^2 \ll m_1^2$, and with the momentum transfer ${\bf q}:={\bf p}_1-{\bf p}_3$,\[\renewcommand{\arraystretch}{1.25} \begin{array}{rl}{\ds \frac{1}{(2\pi)^3} \int} d{\bf x}\; e^{i{\bf q}\cdot {\bf x}}\;V({\bf x})&= i {\cal M}(({\bf p},w)_4)\\
 &= {\ds \frac{(2\pi)^3 c_4}{m^2}}\;\overline{U_s(0, 2{\bf p}_1)}\,U_s(0,2{\bf p}_1\!-\!2{\bf q}).\end{array}\]is one relation between $V(x)$ and ${\cal M}(({\bf p},w)_4)$ that results in equal cross sections. Equivalent potentials $V(x)$ that depend on the energy of the collision, given by the incident momentum ${\bf p}_1$ in this center of momentum frame, are a general feature of the constructions. But, the Yukawa and $1/r$ potentials are examples of potentials that are nearly independent of the incident momentum. A relative phase is arbitrary when cross sections rather than amplitudes are equated. Fourier transformation provides that\[e^{i{\bf p}_1\cdot {\bf x}} \;V({\bf x})= \int d{\bf q}\; e^{i({\bf p}_1-{\bf q})\cdot {\bf x}}  \frac{(2\pi)^3 c_4}{m^2}\;\overline{U_s(0, 2{\bf p}_1)}\,U_s(0,2{\bf p}_1\!-\!2{\bf q})\]or\begin{equation}\label{eff-pot}V({\bf x})= e^{-i{\bf p}_1\cdot {\bf x}}\,\frac{(2\pi)^3 c_4}{m^2}\;\overline{U_s(0, 2{\bf p}_1)}\,\;\int d{\bf q}'\; e^{i{\bf q}'\cdot {\bf x}} U_s(0,2{\bf q}').\end{equation}For macroscopic distances and typical momenta, ${\bf p}_1\cdot {\bf x}\gg 1$ and is unobservable in a finite-sized detector. Rotational invariance is exploited to align axes for the summation with the result that\[\renewcommand{\arraystretch}{2} \begin{array}{rl} {\ds \int} d{\bf q} \; e^{i{\bf q}\cdot{\bf x}} \varphi(q) &= {\ds \int_0^\infty} q^2dq \;\varphi(q) {\ds \int_0^\pi} \sin \theta d\theta \; e^{iqr\cos \theta} {\ds \int_0^{2\pi}} d\phi\\
&= {\ds \frac{4\pi}{r}}{\ds \int_0^\infty} dq \;q\,\varphi(q)\sin(qr)\end{array}\]with $q^2:= \|{\bf q}\|^2$ and $r^2:=\|{\bf x}\|^2$.

A Yukawa potential results for the square summable, infinitely differentiable\[U_s(0,2{\bf q})= \frac{1}{1+\delta^2 q^2}+ \frac{\alpha}{\delta^2\,(\epsilon^2+q^2)}\]described with an arbitrarily small length $\delta>0$, momentum parameter $\epsilon>0$, and constant $\alpha>0$. This $U_s(0,2{\bf p}_1)$ is nearly constant for $q \ll 1/\delta$ and\[\frac{1}{1+(\delta q)^2}\gg \frac{\alpha}{\delta^2\,(\epsilon^2+q^2)}.\]This is satisfied for $\alpha \ll \delta^2\epsilon^2$ and $\alpha\leq 1$. The Fourier sine transform is\[{\ds \int} d{\bf q} \; e^{i{\bf q}\cdot{\bf x}} U_s(0,2{\bf q}) = {\ds \frac{2\pi^2}{\delta^2 r}} \left(e^{-r/\delta}+\alpha\,e^{-\epsilon r}\right)\]results from\[\renewcommand{\arraystretch}{2.25} \begin{array}{rl} {\ds \int_0^\infty}dq \;{\ds \frac{q\,\sin(qr)}{a^2+q^2}} &= {\ds \frac{1}{2i}}{\ds \int_{-\infty}^\infty}dq \;{\ds \frac{q\,e^{iqr}}{a^2+q^2}}\\
 &= {\ds \frac{\pi}{2}} \, e^{-ra}\end{array}\]due to the Cauchy integral and Plancherel theorems when $r>0$. The maximum energy that satisfies the conditions is set by $q \ll 1/\delta$ and $\delta$ may be arbitrarily small.

The potential (\ref{eff-pot}) is in the form of a Yukawa potential\[V({\bf x})\approx C\, \frac{e^{-\epsilon r}}{r}\]when\[e^{-r/\delta}\ll \alpha\,e^{-\epsilon r}.\]For $\alpha< \epsilon^2 \delta^2 <1$, this applies when $r\gg -\delta\,\ln\alpha/(1-\epsilon\delta)>0$. Inside a minimum distance roughly proportional to $\delta$, the potential will deviate from $e^{-\epsilon r}/r$. Selection of the coefficient $c_4$ sets the strength of the potential, $|C|=16\pi^5\,\alpha\,c_4/(m \delta)^2$.

A $1/r$ potential results for $-\delta\,\ln\alpha/(1-\epsilon\delta) \ll r \ll 1/\epsilon$ and $\epsilon>0$ may be arbitrarily small. $\epsilon=0$ is excluded for $U_2(p)$ to be a multiplier. Without further development of the classical limits for the constructions, $\epsilon$ is not associated with an elementary particle's mass. In the designations of this note, the elementary particle masses $m_\kappa$ appear in the two-point function.

When the Heaviside function is mollified to smoothly exclude a neighborhood of the point $p^2=0$, a Poincar\'{e} invariant selection for $U_s(2p)$ is\[U_s(2p)= \left(\frac{1}{1+\delta^2 p^2}+ \frac{\alpha}{\delta^2\,(\epsilon^2+p^2)}\right)\, \theta(p^2).\]

% = = = = = = Electrodynamics model = = = = = =
\section{Compton Scattering}

Compton scattering is used to demonstrate that a Feynman rules scattering amplitude is within the range of the constructions. It is demonstrated that the constructive method can approximate first contributing order, two-in two-out scattering amplitudes from Feynman series. The demonstration establishes that quantum mechanical models can approximate the phenomenologically justified methods of Feynman series.

The contrast is made in three dimensional space, $d=4$. The result is that cross sections agree, and scattering amplitudes agree up to a phase, for small momentum transfers and for transverse photon polarizations. This polarization condition is non-covariant, the result of the noncovariant Coulomb condition with the covariant Lorentz condition. Satisfaction of the polarization constraints is enabled in electrodynamics by the gauge invariance of Maxwell's equations. The construction approximating quantum electrodynamics is anticipated to be described in classical limits by fields satisfying Maxwell's equations and these limits exhibit gauge invariance. In the constructions, this exclusion of polarization states suffices to have a semi-norm.

Furry's theorem [\ref{weinberg}] provides that the odd order functions vanish in a Feynman rules development of electrodynamics. This result from QED is naturally exhibited by the constructions. The constructions have vanishing odd-point functions.

%= = = = = = = = = = = = = = = = = = = = = = =
\subsection{Feynman rules and VEV}

The Compton scattering amplitude [\ref{weinberg},\ref{schwabl}] is\begin{equation}\label{compton-sa}\langle a^*_{\mathit{out},1} b^*_{\mathit{out},2}\Omega|a^*_{\mathit{in},3}b^*_{\mathit{in},4}\Omega\rangle = 2\pi i\,\delta(p_1\!+\!p_2\!-\!p_3\!-\!p_4) {\cal M}\end{equation}with\[ \renewcommand{\arraystretch}{2.25} \begin{array}{rl} {\cal M}&= g_{r_1r_1}g_{r_3r_3}\,{\ds \frac{e^2}{(2\pi)^3}}\,\overline{u}_{r_2}\gamma_0({\not \! \epsilon}_{r_3} {\ds \frac{{\not \! p}_4\!-\!{\not \!p}_1+m}{(p_4\!-\!p_1)^2\!-\!m^2}} {\not \! \overline{\epsilon}}_{r_1}+{\not \! \overline{\epsilon}}_{r_1}{\ds \frac{{\not \! p}_1\!+\!{\not \!p}_2+m}{(p_1\!+\!p_2)^2\!-\!m^2}} {\not \! \epsilon}_{r_3})u_{r_4}\\
 &= g_{r_1r_1}g_{r_3r_3}\, {\ds \frac{e^2}{(2\pi)^3}}\,{\ds \sum_{\alpha_1,\beta_1,\mu_1}}\,(\gamma_0\gamma_{\mu_1}^*)_{\alpha_1\beta_1}u_{r_4,(\beta_1)} {\ds \sum_{\alpha_2,\beta_2,\mu_2}} \,(\gamma_0\gamma_{\mu_2}^*)_{\alpha_2\beta_2} \overline{u}_{r_2,(\alpha_2)}\times\\
 &\quad \left( \overline{\epsilon}_{r_1;(\mu_1)} \epsilon_{r_3;(\mu_2)}{\ds \frac{(({\not \! p}_4\!-\!{\not \! p}_1\!+\!m)\gamma_0)_{\beta_2 \alpha_1}}{(p_4\!-\!p_1)^2\!-\!m^2}}+ \overline{\epsilon}_{r_1;(\mu_2)} \epsilon_{r_3;(\mu_1)}{\ds \frac{(({\not \! p}_1\!+\!{\not \! p}_2\!+\!m)\gamma_0)_{\beta_2 \alpha_1}}{(p_1\!+\!p_2)^2\!-\!m^2}}\right),\end{array}\]to first contributing order. The free field photon creation operators $a_k^*$ from (\ref{photon-c}) create photon states with polarizations $\epsilon_{r_k}(p_k)$ and energy-momenta $p_k$ with $k=1,3$, and the electron free field creation operators $b_k^*$ from (\ref{electron-c}) create electron states with polarizations $w_{p,r_k}(p_k)$ and energy-momenta $p_k$ with $k=2,4$. The energy-momenta $(p)_4$ correspond to $k,p,k',p'$ [\ref{weinberg},\ref{schwabl}]. (\ref{compton-sa}) results from (\ref{instates}), Appendix \ref{app-electro}, and the interaction Hamiltonian density for electrodynamics,\[\renewcommand{\arraystretch}{1.25} \begin{array}{rl} H_I(x) &=e\;:\!\overline{\Psi} (x) \not \!\!A(x) \Psi(x) \!:\\
 &:=e {\ds \sum_{\alpha,\beta,\mu=0}^3} :\!\Psi(x)^*_\alpha \,(\gamma_0 \gamma_\mu^*)_{\alpha \beta}A(x)_\mu \Psi(x)_\beta \!:.\end{array}\]$e$ is the charge. Since photons with time-like polarization components are not observed, there is no contribution from $r_1,r_3=0$ and the two factors of $g_{r_jr_j}=-1$ can be neglected. With the change in notation from (\ref{slash}) to\begin{equation}\label{R-defn}R(p,m):= ({\not \! p}+m)\gamma_0\end{equation}and\[R_m(p):= {\ds \frac{({\not \! p}\!+\!m)\gamma_0}{p^2\!-\!m^2}}= {\ds \frac{R(p,m)}{p^2\!-\!m^2}}\]for $p^2\neq 0$, (\ref{compton-sa}) equals\[{\cal M}= {\ds \frac{e^2}{(2\pi)^3}}\,\overline{u}_{r_2}\gamma_0({\not \! \epsilon}_{r_3} R_m(p_4\!-\!p_1) \gamma_0{\not \! \overline{\epsilon}}_{r_1}+ {\not \! \overline{\epsilon}}_{r_1}R_m(p_1\!+\!p_2) \gamma_0 {\not \! \epsilon}_{r_3})u_{r_4}\]in terms of 4x4 Hermitian matrices $R_m(p)$ and $\gamma_0\gamma_\mu^*$. With energy-momenta on mass shells, $p^2=m^2$ and $q^2=0$ in the Compton scattering case,\begin{equation}\label{p-signs}\renewcommand{\arraystretch}{1.25} \begin{array}{l} (p+ q)^2 -m^2=2pq >0\\
(p- q)^2-m^2 =-2pq <0\end{array}\end{equation}for $m>0$ unless $q_{(0)}=0$ in the rest frame of the fermion and then both equal zero. $q_{(0)}>0$ except for the uninteresting case of infinite wavelength photons.

The Feynman series scattering amplitude (\ref{compton-sa}) derives from VEV evaluated using the Fourier transforms of free fields expressed as creation and annihilation operators, (\ref{photon-f}), (\ref{dirac-f}) and (\ref{dirac-fadj}) from Appendix \ref{app-electro}. Commutation of the free fields $A(x)$ and $\Psi(x)$ and the properties of the free field vacuum state result in the Fourier transform of the four-point function,\[\renewcommand{\arraystretch}{1.25} \begin{array}{l}((D)_2 \cdot {^C\tilde{W}}_4(-\xi_2,-\xi_1,\xi_3,\xi_4)) = \langle \Omega|\tilde{\Psi}(-p_2)_{\kappa_2} \tilde{A}(-p_1)_{\kappa_1} \tilde{A}(p_3)_{\kappa_3} \tilde{\Psi}(p_4)^*_{\kappa_4} \Omega\rangle\\
 \qquad = (2\pi)^2 {\ds \prod_{j=1}^4} \delta(E_j-\omega_j) {\ds \sum_{r_1=1}^4 \epsilon_{r_1,\kappa_1} \sum_{r_2=1}^2 u_{r_2,\kappa_2}\sum_{r_3=1}^4 \overline{\epsilon}_{r_3,\kappa_3} \sum_{r_4=1}^2} \overline{u}_{r_4,\kappa_4} \langle \Omega|a_{\mathit{out},1} b_{\mathit{out},2} a^*_{\mathit{in},3}b^*_{\mathit{in},4}\Omega\rangle. \end{array}\]Discussed in note \ref{foot-notatn} of the Introduction, factors of $D$ result for free field operator-valued distribution adjoints for the $*$-mapped function arguments $p_1,p_2$. Substitution of (\ref{compton-sa}), neglecting the forward contributions, and identities from the photon two-point function (\ref{2pt-photon}),\[\sum_{r=1}^4 g_{rr} \epsilon_r(p)_\alpha \overline{\epsilon}_r(p)_\beta = \frac{g_{\alpha, \beta}}{2\omega}\]and the fermion two-point function (\ref{2pt-PP*}),\[\sum_{r=1}^2 u_r(p)_\alpha \overline{u_r}(p)_\beta = \frac{(({\not \! p}+m)\gamma_0)_{\alpha \beta}}{2\omega}\]result in the connected contribution to the Fourier transform of the four-point function from the first contributing order of the Feynman series for Compton scattering.\begin{equation}\label{feyn-4pt}((D)_2 \cdot {^C\tilde{W}}_4(-\xi_2,-\xi_1,\xi_3,\xi_4)) = \delta^+_1\delta^+_2\delta^+_3\delta^+_4\delta(p_1\!+\!p_2\!-\!p_3\!-\!p_4) \;{\cal V}\end{equation}with\[\renewcommand{\arraystretch}{1.75} \begin{array}{rl} {\cal V} &= ie^2\,{\ds \sum_{\alpha_1,\beta_1,\mu_1}}(\gamma_0\gamma_{\mu_1})_{\alpha_1\beta_1}(({\not \! p}_4 +m)\gamma_0)_{\beta_1 \kappa_4} {\ds \sum_{\alpha_2,\beta_2,\mu_2}}(\gamma_0\gamma_{\mu_2})_{\alpha_2\beta_2} (({\not \! p}_2+m)\gamma_0)_{\kappa_2 \alpha_2}\times\\
 & \qquad g_{\mu_1\mu_1} g_{\mu_2\mu_2}\;\left( g_{\mu_1\kappa_1} g_{\mu_2\kappa_3} {\ds \frac{(({\not \! p}_4\!-\!{\not \! p}_1\!+\!m)\gamma_0)_{\beta_2 \alpha_1}}{(p_4\!-\!p_1)^2\!-\!m^2}}+ g_{\mu_2\kappa_1} g_{\mu_1\kappa_3} {\ds \frac{(({\not \! p}_1\!+\!{\not \! p}_2\!+\!m)\gamma_0)_{\beta_2 \alpha_1}}{(p_1\!+\!p_2)^2\!-\!m^2}}\right)\\
 &:= {\cal V}_s((\xi)_4)+{\cal V}_u((\xi)_4)\end{array}\]and\begin{equation}\label{m-expnd}\renewcommand{\arraystretch}{1.5} \begin{array}{rl} {\cal V}_s((\xi)_4) &:= a_s((p)_4)\left( ({\not \! p}_2 +m) \gamma_{\kappa_1}R(p_1\!+\!p_2,m) \gamma_{\kappa_3}^* ({\not \! p}_4+m)^*\right)_{\kappa_2 \kappa_4}\\
{\cal V}_u((\xi)_4) &:= a_u((p)_4)\left( ({\not \! p}_2 +m) \gamma_{\kappa_3} R(p_4\!-\!p_1,m)\gamma_{\kappa_1}^* ({\not \! p}_4+m)^*\right)_{\kappa_2 \kappa_4}.\end{array}\end{equation}This result uses (\ref{gamma-star}), $g_{\mu\mu}g_{\mu\kappa}=\delta_{\mu\kappa}$, $\gamma_0^2=1$ and $2\omega_k\, \delta^+_k:=\delta(E_k-\omega_k)$. The Feynman rules result is\begin{equation}\label{feyn-a}a_s((p)_4)=\frac{ie^2}{(p_1\!+\!p_2)^2\!-\!m^2} \qquad \mbox{and} \qquad a_u((p)_4)=\frac{ie^2}{(p_4\!-\!p_1)^2\!-\!m^2}.\end{equation}

To study definiteness of the VEV, a generalization of (\ref{feyn-4pt}) is introduced. The generalization of the connected four-point function (\ref{feyn-4pt}) is a substitution of multiplier functions $a_s(p),a_u(p)$ for the Feynman rules coefficients (\ref{feyn-a}). The first contributing order, Feynman rules connected four-point VEV is imaginary, $i$ times real functions, Hermitian matrices and delta functions. For Lorentz invariant multiplier functions $a_s,a_u$, the generalization has the same Poincar\'{e} covariance as the Feynman rules result (\ref{feyn-4pt}) with (\ref{feyn-a}).

The four-point VEV derived from the Feynman rules is Lorentz covariant. Together with the nonnegativity developed in the following section, the Lorentz covariance demonstrates that an approximation of the Feynman rules result satisfies the revised Wightman axioms. Satisfaction of the locality and spectral support conditions follows from the symmetrization (\ref{permutation}) and energy support (\ref{feyn-4pt}). Lorentz covariance derives from the transformation of the field $\Phi(\Lambda^{-1}x)=S(A)\Phi(x)$ that in field components (\ref{electro-field}) is\[A(\Lambda^{-1}x) = \Lambda(A)^{-1} A(x), \qquad \Psi(\Lambda^{-1}x)=\overline{S}_p(A)\Psi(x), \qquad \Psi(\Lambda^{-1}x)^*=S_p(A) \Psi(x)^*\]from (\ref{lorentz1}) and (\ref{lorentz-fermion-full}). In a Poincar\'{e} transformed coordinate frame, the VEV are\[\renewcommand{\arraystretch}{1.25} \begin{array}{l}\langle \Omega|\tilde{\Psi}(-\Lambda^{-1}p_2)_{\kappa_2} \tilde{A}(-\Lambda^{-1}p_1)_{\kappa_1} \tilde{A}(\Lambda^{-1}p_3)_{\kappa_3} \tilde{\Psi}(\Lambda^{-1}p_4)^*_{\kappa_4} \Omega\rangle = ie^2\,\delta(p_1\!+\!p_2\!-\!p_3\!-\!p_4) \prod_j\; \delta^+_j \times\\
 \qquad\qquad\qquad\qquad \left((\Lambda^{-1}{\not \! p}_2 +m) \gamma_{\kappa_3} R_m(\Lambda^{-1}(p_4\!-\!p_1)) \gamma_0\gamma_{\kappa_1} (\Lambda^{-1}{\not \! p}_4+m)\gamma_0\right.\\
 \qquad\qquad\qquad\qquad\quad \left.+ (\Lambda^{-1}{\not \! p}_2 +m) \gamma_{\kappa_1} R_m(\Lambda^{-1}(p_1\!+\!p_2)) \gamma_0\gamma_{\kappa_3} (\Lambda^{-1}{\not \! p}_4+m)\gamma_0 \right)_{\kappa_2 \kappa_4}\end{array}\]from (\ref{feyn-4pt}) and (\ref{m-expnd}) using the Lorentz invariance of $p=0$, $p^2$ and $E>0$ for proper orthochronous Lorentz transformations. $\Lambda^{-1}{\not \! p}:=R(\Lambda^{-1}p,0)\gamma_0$ using (\ref{R-defn}). The realization of (\ref{matcond2}) for the electron, (\ref{lorentz-fermion-full}), results in (\ref{sp-iden2}), (\ref{sp-iden}) and\[\overline{S_p}(A)({\not \!p}+m)\gamma_0 S_p(A)^T =(\Lambda^{-1}{\not \!p}+m)\gamma_0.\]For the first term from (\ref{m-expnd}), this identity and (\ref{R-defn}) result in\[\renewcommand{\arraystretch}{1.25} \begin{array}{l} \left( (\Lambda^{-1}{\not \! p}_2 +m) \gamma_{\kappa_3} R(\Lambda^{-1}(p_4\!-\!p_1)) \gamma_0 \gamma_{\kappa_1} (\Lambda^{-1}{\not \! p}_4+m)\gamma_0 \right)_{\kappa_2 \kappa_4}\\
 \qquad =\left(\overline{S_p}({\not \! p}_2 +m)\gamma_0 S_p^T \gamma_0 \gamma_{\kappa_3} \overline{S_p} R(p_4\!-\!p_1) S_p^T \gamma_0 \gamma_{\kappa_1} \overline{S_p}({\not \! p}_4+m) \gamma_0 S_p^T\right)_{\kappa_2 \kappa_4}\\
 \qquad = \left( \overline{S_p}({\not \! p}_2 +m)\gamma_0 S_p^T \gamma_{\kappa_3}^* \gamma_0 \overline{S_p} R(p_4\!-\!p_1) S_p^T \gamma_{\kappa_1}^*\gamma_0 \overline{S_p}({\not \! p}_4+m)\gamma_0 S_p^T\right)_{\kappa_2 \kappa_4}\end{array}\]using $\gamma_0^2=1$ and (\ref{gamma-star}). A similar result applies for the second term in (\ref{m-expnd}). (\ref{sp-iden2}) and (\ref{sp-iden}) (cf. equation 1-43 [\ref{pct}]) provides that\[S_p(A)^T \gamma_\kappa^* \gamma_0 \overline{S_p}(A)= \sum_\nu \Lambda_{\kappa \nu}^{-1} \gamma_\nu^* \gamma_0.\]Together with Lorentz invariance of the coefficients $a_s$ and $a_u$, this demonstrates the Poincar\'{e} covariance of (\ref{feyn-4pt}). That is,\[\renewcommand{\arraystretch}{1.25} \begin{array}{l}\langle \Omega|\tilde{\Psi}(-\Lambda^{-1}p_2)_{\kappa_2} \tilde{A}(-\Lambda^{-1}p_1)_{\kappa_1} \tilde{A}(\Lambda^{-1}p_3)_{\kappa_3} \tilde{\Psi}(\Lambda^{-1}p_4)^*_{\kappa_4} \Omega\rangle\\
 \qquad =\langle \Omega|(\overline{S}_p \tilde{\Psi}(-p_2))_{\kappa_2} (\Lambda^{-1}\tilde{A}(-p_1))_{\kappa_1} (\Lambda^{-1} \tilde{A}(p_3))_{\kappa_3} (S_p \tilde{\Psi}(p_4)^*)_{\kappa_4} \Omega\rangle.\end{array}\]

\subsection{Nonnegativity of the ${^C\!W}_4$ approximating the Feynman series}

Variations of the constructions are required to approximate the first contributing order from Feynman series for Compton scattering. While the constructions result in field theories that exhibit interaction with desired Poincar\'{e} covariance properties, (\ref{m-expnd}) is not proportional to a selection for (\ref{scatt-eval}). Variations of the constructions include additional organizations of the two-point function $M(p)$, submatrices, and Lorentz invariant functions into connected functions.

In this section, a connected four-point function that approximates the Feynman series Compton scattering cross section is developed. (\ref{feyn-4pt}) consists of the two terms (\ref{m-expnd}). The constructions replicate the ${\cal V}_s$ term and approximate the ${\cal V}_u$ term at small momentum transfers. The $a_u$ that provides nonnegativity has a fractional error from the Feynman series result (\ref{feyn-4pt}) proportional to the momentum exchange $(p_4-p_2)/m$. 

The constructed connected four-point functionals are nonnegative.\begin{equation}\label{4pt-trun} {^C\!W}_4(f_2^* \,{\bf x}\, f_2)=\int d(\xi)_4\; ((D)_2 \cdot {^C\tilde{W}}_4(-\xi_2,-\xi_1,\xi_3,\xi_4))\, \overline{\tilde{f}_2}(\xi_1,\xi_2)\tilde{f}_2(\xi_3,\xi_4) \geq 0.\end{equation}This nonnegativity results in the demonstration of a semi-norm (\ref{norm}) for the constructions [\ref{gej05}]. The Feynman rules result in an indefinite connected four-point function but the approximation times $-i$ is nonnegative in the subspace of transversely polarized photon states.

%= = = = = = = = = = = = = = = = = = = = =
%      P1 condition
%= = = = = = = = = = = = = = = = = = = = =
To achieve the nonnegativity (\ref{4pt-trun}), it is sufficient that the constructed VEV (\ref{feyn-4pt}) be a summation of terms in the form:\begin{equation}\label{p1}{\cal V}_a((\xi)_4) := {\ds \int} \mu_a(dv)\; \overline{T_a(v,\xi_1,\xi_2)} T_a(v,\xi_3,\xi_4)\end{equation}with $v$ a set of values summed using nonnegative measure $\mu_a(dv)$. This factorization applies in the subspace of transversely polarized photon states (\ref{pol-st-photon}), for energies $E_j$ on the positive mass shells, and on the manifold with energy-momentum conservation, $p_1+p_2=p_3+p_4$. Transversely polarized here indicates satisfaction of both the Coulomb, $w_\epsilon(p)_0=0$, and Lorentz, $p\,w_\epsilon(p)=0$, conditions with $p$ the photon energy-momentum.

The sufficiency of (\ref{p1}) is demonstrated by substitution into (\ref{feyn-4pt}) and (\ref{4pt-trun}) using that sums of nonnegative numbers are nonnegative.\[{^C\!W}_4(f_2^* \,{\bf x}\, f_2) = \sum_a \int \frac{du}{(2\pi)^d} \int \mu_a(dv)\; \left|{\ds \int}d(\xi)_2\;e^{iu(p_1\!+\!p_2)}\delta_1^+\delta_2^+\,T_a(v,(\xi)_2) f_2((\xi)_2)\right|^2\geq 0\]from\[(2\pi)^d \delta(p)=\int du\; e^{ipu}.\]$u$ is a $d$-dimensional spacetime vector. Following (\ref{dualf}) and (\ref{electro-field}), the functions $\tilde{f}_2(p_1,p_2)_{\kappa_1 \kappa_2}$ appropriate for Compton scattering have $0\leq \kappa_1 \leq 3$ and $8\leq \kappa_2 \leq 11$.

To achieve (\ref{p1}) with a four-point VEV of the form (\ref{feyn-4pt}) with (\ref{m-expnd}), it is sufficient that:\begin{enumerate} \item $R(p_1\!+\!p_2,m)$ is a positive semidefinite 4x4 matrix,
\item Summed over $\kappa_1$ and $\kappa_3$ for photon states with polarizations that satisfy the Coulomb and Lorentz conditions,\[({\not \! p}_2\!+\!m)\gamma_{\kappa_3}R(p_4\!-\!p_1,m)\gamma_{\kappa_1}^*({\not \! p}_4\!+\!m)^*=({\not \! p}_2\!+\!m)\gamma_{\kappa_3}R(-p_2\!-\!p_1,m)\gamma_{\kappa_1}^*({\not \! p}_4\!+\!m)^*\]and,
\item $[\gamma_{\kappa_3}R(-p_2\!-\!p_1,m)\gamma_{\kappa_1}^*]$ is a negative semidefinite 16x16 matrix.\end{enumerate}$[\gamma_{\kappa_3}R(-p_2\!-\!p_1,m)\gamma_{\kappa_1}^*]$ is the 16x16 matrix described by the 4x4 arrangement of 4x4 submatrices $\gamma_{\kappa_3}R(-p_2\!-\!p_1,m)\gamma_{\kappa_1}^*$. 

% = = = = = Sufficiency = = = = = =
Sufficiency of the conditions is demonstrated first, and then the validity of the assertions is established for an approximation to the Feynman series result for Compton scattering. Validity is first established in particular, convenient frames of reference and finally the demonstrations are extended to appropriate, general frames.

The asserted positive semidefiniteness of the matrices results in factorizations, $M=C^*C$ [\ref{horn}]. From energy-momentum conservation, $p_1+p_2=p_3+p_4$,\[\renewcommand{\arraystretch}{1.25}\begin{array}{rl} R(p_1+p_2,m) &=C_e^*(p_1+p_2,m) C_e(p_1+p_2,m)\\
 &=C_e^*(p_1+p_2,m) C_e(p_3+p_4,m)\end{array}\]and,\[-[\gamma_{\kappa_3}R(-p_2\!-\!p_1,m)\gamma_{\kappa_1}^*]=C_x(p_1\!+\!p_2)^*C_x(p_3\!+\!p_4).\]With $a_s,a_u$ equal to $-i$ times (\ref{feyn-a}), $a_s>0$ and $a_u<0$ from (\ref{p-signs}). The nonnegative root of $-a_u$ is not appropriate for (\ref{p1}), but\begin{equation}\label{constrt-a}a_u((p)_4)=\frac{e^2}{(p_2\!-\!p_1)^2\!-\!m^2}=\frac{e^2}{2p_1 p_2}\end{equation}has the proper form. Energy-momentum conservation establishes that $p_1p_2=p_3p_4$. The substitution approximates the cross section derived from (\ref{feyn-a}) when momentum exchanges are small, ${\bf p}_4 - {\bf p}_2\approx 0$. (\ref{m-expnd}) with $a_s$ equal to $-i$ times the Feynman rules $a_s$ (\ref{feyn-a}) and $a_u$ from (\ref{constrt-a}) have decompositions that satisfy (\ref{p1}). These decompositions are\[\renewcommand{\arraystretch}{1.25}\begin{array}{rl}  T_s(v,\xi_1,\xi_2) &= \sqrt{a_s}\,(C_e(p_1+p_2,m)\gamma_{\kappa_1}^* ({\not \! p}_2+m)^*)_{\ell,\kappa_2}\\
 T_u(v,\xi_1,\xi_2) &= \sqrt{-a_u}\;{\ds \sum_{j=0}^3}\, C_x(p_1+p_2)_{\ell,\kappa_1j} ({\not \! p}_2+m)^*_{j,\kappa_2}\end{array}\]with $v=\ell$ summed from 1 to 4 for $T_s$ and from 1 to 16 for $T_u$, and both $a_s$ and $a_u$ are functions of $p_1p_2=p_3p_4$ and the semidefiniteness, from (\ref{p-signs}), is appropriate.

% = = = = = Positive semidefiniteness of $R(p_1\!+\!p_2,m)$ = = = = = =
Validity of the three assertions is now developed. $R(p_1\!+\!p_2,m)$ is positive semidefinite. From [\ref{horn}, theorem 7.2.7], positive semidefiniteness is demonstrated given a factorization, $M=C^*C$. For a general $R(p,\mu)$ from (\ref{R-defn}) with $p^2\geq \mu^2$, (\ref{slash}), (\ref{gamma}) and (\ref{P-fermion}), $P(p)^*=P(p)$, and $P(0,{\bf p})^2={\bf p}^2$ proportional to the identity matrix provide that\begin{equation}\label{slash-factor}\renewcommand{\arraystretch}{1.75} \begin{array}{rl} R(p,\mu) &=\left(\renewcommand{\arraystretch}{1}\begin{array}{cc} (E+\mu)\sigma_0 & P(0,{\bf p})\\ P(0,{\bf p}) & (E-\mu)\sigma_0\end{array}\right)\\
 &= {\ds \frac{1}{E+\mu}}\left(\renewcommand{\arraystretch}{1}\begin{array}{cc} (E+\mu)\, \sigma_0 & 0\\ P(0,{\bf p})^* & \sqrt{p^2\!-\!\mu^2}\,\sigma_0\end{array}\right)\left(\renewcommand{\arraystretch}{1}\begin{array}{cc} (E+\mu)\, \sigma_0 & P(0,{\bf p}) \\ 0 & \sqrt{p^2\!-\!\mu^2}\,\sigma_0\end{array}\right)\\
 &:= C_e^*(p,\mu) C_e(p,\mu)\end{array}\end{equation}when $E\geq 0$. Indeed, the doubly redundant eigenvalues of $R(p,\mu)$ are nonnegative, two sets of $\lambda_\pm=E\pm \sqrt{\mu^2+{\bf p}^2}\geq 0$. $\mu$ may be positive or negative. The case of interest is $p=p_1+p_2$ and $\mu=m$.

% = = = = = Identities among energy-momenta dependencies for $R(p_4\!-\!p_1,m)$ = = = = = =
The identities among the $R(\pm p_j-p_1,m)$ for $j=2,4$ in the context of (\ref{m-expnd}) appear in evaluations of spin averaged cross sections [\ref{weinberg}]. The representation for the covering group of Lorentz transformations and that observed photon states satisfy the Coulomb and Lorentz conditions result in the identities. Summed over $\kappa_3$ for any photon polarization state that satisfies the Coulomb condition,\[\renewcommand{\arraystretch}{1.25} \begin{array}{rl}{\not \! w}_\epsilon(p) &:= {\ds \sum_{\kappa_3}} \gamma_{\kappa_3}^* w_\epsilon(p)_{\kappa_3}\\
 &=\left(\renewcommand{\arraystretch}{1.25} \begin{array}{cc}0& -P(0,{\bf w}_\epsilon)\\ P(0,{\bf w}_\epsilon)& 0\end{array}\right)\end{array}\]from (\ref{slash}), (\ref{P-fermion}) and $w_\epsilon=(0,{\bf w}_\epsilon)$. In the rest frame of the product electron, here labeled $p_4=(m,0,0,0)$,\[\renewcommand{\arraystretch}{1.25} \begin{array}{rl}({\not \!p}_4\!+\!m){\not \! w}_\epsilon(p)({\not \!p}_4\!+\!m)^*&=\left(\renewcommand{\arraystretch}{1.25} \begin{array}{cc} 2m& 0\\ 0& 0\end{array}\right)\left(\renewcommand{\arraystretch}{1.25} \begin{array}{cc}0& -P(0,{\bf w}_\epsilon)\\ P(0,{\bf w}_\epsilon)& 0\end{array}\right)\left(\renewcommand{\arraystretch}{1.25} \begin{array}{cc} 2m& 0\\ 0& 0\end{array}\right)\\
 &=0\end{array}\]independently of the photon momentum $p$. ${\not \! w}_\epsilon$ covers all possibilities for factors of $\gamma_{\kappa_3}^*$ allowed by observed photon states. Then, using the notation $R(p,m)$ from (\ref{R-defn}), the contribution of $p_4$ to $R(p_4-p_1,m)$ vanishes due to the multiplication by $({\not \! p}_4\!+\!m)^*$.\begin{equation}\label{Rl-part}R(p_4\!-\!p_1,m) \gamma_{\kappa_3}^* ({\not \! p}_4+m)^* =R(-p_1,0) \gamma_{\kappa_3}^* ({\not \! p}_4+m)^*\end{equation}when photon polarizations are restricted to satisfy the Coulomb condition. In the rest frame of the incident electron labeled $p_2$, the adjoint of (\ref{Rl-part}) provides that\[({\not \! p}_2+m) \gamma_{\kappa_1}R(-p_2\!-\!p_1,m) =({\not \! p}_2+m) \gamma_{\kappa_1}R(-p_1,0).\]Substitution results in the desired identity.\begin{equation}\label{Rl-ident}\renewcommand{\arraystretch}{1.25} \begin{array}{rl} ({\not \! p}_2+m) \gamma_{\kappa_1} R(-p_1,0) \gamma_{\kappa_3}^* ({\not \! p}_4+m)^* &=({\not \! p}_2+m) \gamma_{\kappa_1} R(p_4\!-\!p_1,m) \gamma_{\kappa_3}^* ({\not \! p}_4+m)^*\\
 &=  ({\not \! p}_2+m) \gamma_{\kappa_1} R(-p_2\!-\!p_1,m) \gamma_{\kappa_3}^* ({\not \! p}_4+m)^*
\end{array}\end{equation}when photon polarizations are restricted to satisfy the Coulomb and Lorentz conditions. Validity of the identity for all appropriate values of $p_2$ and $p_4$ requires the Lorentz condition in addition to the Coulomb condition, as discussed below.

% = = = = = Negative definiteness of $\gamma_{\kappa_3}R(-p_2\!-\!p_1,m)\gamma_{\kappa_1}^*$ = = = = =
$[\gamma_{\kappa_3} R(-p_2-p_1) \gamma_{\kappa_1}^*]$ is a negative semidefinite 16x16 matrix. In the rest frame of the incident electron labeled by $p_2=(m,0,0,0)$, a reference frame achievable through a rotation has $p_1=(\varrho_1,0,0,\varrho_1)$ with $\varrho_1>0$. In this frame, the Coulomb and Lorentz conditions imply that the photon labeled $p_1$ has polarization states of the form $w_\epsilon=(0,w_{\epsilon(1)},w_{\epsilon(2)},0)$. As a consequence and in this frame, only the 8x8 $\kappa_1,\kappa_3=1,2$ submatrix of $[\gamma_{\kappa_3} R(-p_2-p_1) \gamma_{\kappa_1}^*]$ contributes to the determination of definiteness. The contributing 8x8 principal submatrix can be expressed in 4x4 submatrices as\[-[\gamma_{\kappa_3}R(-p_1\!-\!p_2,m) \gamma_{\kappa_1}^*]= \left(\renewcommand{\arraystretch}{1}\begin{array}{cc} \gamma_1 R(p_1\!+\!p_2,-m) \gamma_1^* & \gamma_2 R(p_1\!+\!p_2,-m) \gamma_1^*\\ \gamma_1 R(p_1\!+\!p_2,-m) \gamma_2^* & \gamma_2 R(p_1\!+\!p_2,-m) \gamma_2^*\end{array}\right)\]using that $-R(-p,m)=R(p,-m)$ from (\ref{R-defn}). Multiplying out the 4x4 matrices using the evaluation of $R(p_1\!+\!p_2,-m)$ from (\ref{slash}) in the selected reference frame, (\ref{gamma}) and (\ref{gamma-star}) provide that\[\renewcommand{\arraystretch}{1.25} \begin{array}{rl}\gamma_i R(p_1\!+\!p_2,-m) \gamma_j^* &=\left(\renewcommand{\arraystretch}{1.25} \begin{array}{cc}0& \sigma_i\\ -\sigma_i & 0\end{array}\right)\left(\renewcommand{\arraystretch}{1.25} \begin{array}{cc}\varrho_1&  \varrho_1 \,\sigma_3\\ \varrho_1 \,\sigma_3  & \varrho_1+2m\end{array}\right)\left(\renewcommand{\arraystretch}{1.25} \begin{array}{cc}0& -\sigma_j\\ \sigma_j & 0\end{array}\right)\\
 &=\left(\renewcommand{\arraystretch}{1.25} \begin{array}{cc}(\varrho_1+2m)\sigma_i\sigma_j& -\varrho_1 \,\sigma_i\sigma_3\sigma_j\\ -\varrho_1 \,\sigma_i\sigma_3\sigma_j  & \varrho_1\sigma_i\sigma_j\end{array}\right)\\
  &=\left(\renewcommand{\arraystretch}{1.25} \begin{array}{cc}\sigma_i\sigma_j & 0\\ 0 & \sigma_i\sigma_j\end{array}\right) \left(\renewcommand{\arraystretch}{1.25} \begin{array}{cc}\varrho_1+2m& -\varrho_1 \,\sigma_j\sigma_3\sigma_j\\ -\varrho_1 \,\sigma_j\sigma_3\sigma_j  & \varrho_1 \end{array}\right)\\
 &=\gamma_i\gamma_j^* R(p_1\!+\!p_2,m).\end{array}\]The result follows from products of the Pauli spin matrices (\ref{paulispin}), that $\sigma_j^2=\sigma_0$ and $\sigma_j\sigma_3\sigma_j=-\sigma_3$ for $j=1,2$. Then,\[\renewcommand{\arraystretch}{1.25} \begin{array}{rl} -[\gamma_{\kappa_3}R(-p_1\!-\!p_2,m) \gamma_{\kappa_1}^*] &= \left(\renewcommand{\arraystretch}{1}\begin{array}{cc} R & i\chi_3\, R\\ -iR\, \chi_3 & R\end{array}\right)\\
 &= \left(\renewcommand{\arraystretch}{1}\begin{array}{cc} C_e^* & 0\\ -iR\,\chi_3C_e^{-1} & 0\end{array}\right)\left(\renewcommand{\arraystretch}{1}\begin{array}{cc} C_e & i(C_e^*)^{-1} \chi_3 \,R\\ 0 & 0\end{array}\right)\\
 &:=C_x(p_1\!+\!p_2)^*C_x(p_3\!+\!p_4)\\\end{array}\]that is positive semidefinite and in the form of (\ref{p1}). $R:=R(p_1\!+\!p_2,-m)$ and $C_e=C_e(p_1\!+\!p_2,-m)$ from (\ref{slash-factor}). The result follows from conservation of energy-momentum, $p_1+p_2=p_3+p_4$, commutation of $R$ with $\chi_3$,\[\chi_3 := -i\gamma_1\gamma_2^* =\left(\renewcommand{\arraystretch}{1}\begin{array}{cc} \sigma_3 & 0\\ 0& \sigma_3 \end{array}\right)\]and $\sigma_1\sigma_2=-\sigma_2\sigma_1=i\sigma_3$. $\chi_3^2=1$ and $\chi_3$ has been designated $-i\Gamma^T_{12}$ and $-i\sigma_{12}$ [\ref{schwabl}]. This factorization completes the demonstration that $[\gamma_{\kappa_3} R(-p_2-p_1) \gamma_{\kappa_1}^*]$ is negative semidefinite in the selected reference frame.

% = = = = = = = = = = = = = = = = = = = = = = = = = = = = = = = = = = = =
% = = = = Lorentz invariance of positive semidefiniteness proved
% = = = = in rest frame of outgoing electron
% = = = = = = = = = = = = = = = = = = = = = = = = = = = = = = = = = = = =
The demonstrations that\[({\not \!p}_k\!+\!m){\not \! w}_\epsilon(p)({\not \!p}_k\!+\!m)=0\]for $k=2,4$ and consequently that $[\gamma_{\kappa_3}R(p_4\!-\!p_1,m)\gamma_{\kappa_1}^*]$ is a negative semidefinite 16x16 matrix in the context of (\ref{m-expnd}) are valid in the rest frames of the electrons $p_4$ and $p_2$ respectively. Validity of the identities and the demonstration of definiteness are extended to all frames by demonstrating Lorentz transformations that preserve the generally noncovariant Coulomb condition and reach all relevant energy-momenta. Lorentz transformation is a $*$-congruence and preserves definiteness.

Every element $A\in$SL(2,C) has a polar decomposition, $A=UP$ with a unitary $U$ and positive semidefinite $P$. Every positive semidefinite $P$ is unitary similar to a diagonal matrix, $P=VDV^*$ [\ref{horn}]. This expresses every $A=UVDV^*$ as rotations $UV$ and $V^*$ and a boost along the $z$-axis $D$ [\ref{bogo}]. Then, with an initial rotation $V^*$, every Lorentz transformation equals a boost along the $z$-axis followed by a rotation, $S(A)=S(UV)S(D)S(V^*)$. In the rest frame of the massive particle, $p_4=(m,0,0,0)$, and considering $V^*$ as the rotation to a reference frame with the photon momentum ${\bf p}_1$ aligned with the $z$-axis, the transverse polarization states are\[w_\epsilon=w_{o\epsilon}:=(0,w_{\epsilon(1)},w_{\epsilon(2)},0)\]and a boost along the $z$-axis followed by a rotation reaches all possibilities for sums of the two energy-momentum vectors attainable by proper, orthochronous Lorentz transformation.\[p_4+\alpha p_1=\left(\renewcommand{\arraystretch}{1} \begin{array}{c} m+\alpha \varrho_1\\0\\0\\ \alpha \varrho_1\end{array}\right)
\mapsto \left(\begin{array}{c} E'\\0\\0\\ \varrho'_1\end{array}\right)
\mapsto \left(\begin{array}{c} E'\\ \sin\phi\,\cos\theta\,\varrho'_1\\ \sin\phi\,\sin\theta\,\varrho'_1\\ \cos\phi\,\varrho'_1\end{array}\right)\]and preserves validity of the Lorentz condition.\[\left(\renewcommand{\arraystretch}{1} \begin{array}{c} 0\\ w_{\epsilon(1)}\\ w_{\epsilon(2)}\\ 0\end{array}\right)
\mapsto \left(\begin{array}{c} 0\\ w_{\epsilon(1)}\\ w_{\epsilon(2)}\\ 0\end{array}\right)
\mapsto \left(\begin{array}{c} 0\\  w_{\epsilon(1)}'\\  w_{\epsilon(2)}'\\ w_{\epsilon(3)}'\end{array}\right).\]The definiteness-implying identity (\ref{Rl-part}) is invariant to the spatial orientation of the photon polarization, that is, is invariant to the intial rotation $V^*$. Consequently, the Coulomb condition is preserved by every Lorentz transformation implemented in the particular sequence $A=UVDV^*$ with the initial polarization $w_\epsilon=S(V^*)^{-1}\,w_\epsilon'$. All such $w_\epsilon$ satisfy the Coulomb and Lorentz conditions when $w_\epsilon'$ does. Consequently, validity of (\ref{Rl-part}) in the particular frames for polarizations that satisfy the Coulomb and Lorentz conditions implies (\ref{Rl-part}) for all $p_4$ and $p_1$ with polarizations that satisfy the Coulomb and Lorentz conditions. Finally, (\ref{Rl-ident}) is valid for all $p_1,p_2,p_4$ and with polarizations that satisfy the Coulomb and Lorentz conditions.

The demonstration of the positive semidefiniteness of the ${\cal V}_s$ term from (\ref{m-expnd}) is independent of the photon polarization, but the demonstration of negative semidefiniteness of the ${\cal V}_u$ term uses the noncovariant Coulomb condition. From (\ref{R-defn}), (\ref{sp-iden2}) and (\ref{sp-iden}) of Appendix \ref{app-eqrep}, the Lorentz transformations of the polarization and energy-momentum derive from\[R(\Lambda^{-1}p,m)= \overline{S_p} R(p,m) S_p^T.\]Then\[\renewcommand{\arraystretch}{1.75} \begin{array}{rl}[\gamma_{\kappa_3}R(\Lambda^{-1}(p_4\!-\!p_1),m)\gamma_{\kappa_1}^*] &=[\gamma_{\kappa_3}\overline{S_p}R(p_4\!-\!p_1,m)S_p^T\gamma_{\kappa_1}^*]\\
 &=[\overline{S_p}\,\overline{S_p}^{-1} \gamma_{\kappa_3} \overline{S_p}R(p_4\!-\!p_1,m)S_p^T\gamma_{\kappa_1}^*(S_p^T)^{-1}S_p^T]\\
 &=[\overline{S_p}\,({\ds \sum_\mu} \Lambda^{-1}_{\kappa_3 \mu} \gamma_\mu)\, R(p_4\!-\!p_1,m)\,({\ds \sum_\nu}\Lambda^{-1}_{\kappa_1 \nu} \gamma_\nu^*)\, S_p^T]\end{array}\]from (\ref{sp-iden}) and its matrix adjoint. The factors of $\Lambda^{-1}$ define the covariantly transformed photon polarization vector. With covariantly redefined polarizations, the Lorentz transform\[ [\gamma_{\kappa_3}R(p_4\!-\!p_1,m)\gamma_{\kappa_1}^*]\mapsto \left(\begin{array}{cccc} \overline{S_p}& 0& 0& 0\\ 0& \overline{S_p}& 0& 0\\ 0& 0& \overline{S_p}& 0\\ 0& 0& 0& \overline{S_p}\end{array}\right) [\gamma_{\kappa_3}R(p_4\!-\!p_1,m)\gamma_{\kappa_1}^*]\left(\begin{array}{cccc}  S_p^T& 0& 0& 0\\ 0& S_p^T& 0& 0\\ 0& 0& S_p^T& 0\\ 0& 0& 0& S_p^T\end{array}\right) \]displays that Lorentz transformation is a $*$-congruence that preserves negative semidefiniteness. It was demonstrated above that the identities (\ref{Rl-ident}) remain valid for all electron and photon energy-momenta with photon polarizations that satisfy the Coulomb condition. The demonstration of negative semidefiniteness follows similar reasoning: every relevant $p_4-p_1$ can be reached by Lorentz transformation of $(m-\varrho_1,0,0,-\varrho_1)$ and preserve validity of the Coulomb condition. $[\gamma_{\kappa_3}R(\Lambda^{-1}(p_4\!-\!p_1),m)\gamma_{\kappa_1}^*]$ depends on only $p_4-p_1$. For every $p_4-p_1$ with $(p_4\!-\!p_1)^2$ constant and sgn$(E_4\!-\!E_1)$ preserved when $(p_4\!-\!p_1)^2\geq 0$, there is a rotation and then a boost along the $z$-axis that puts the matrix in the rest frame of $p_2$ with a photon polarization that satisfies the Coulomb and Lorentz conditions. Consequently, negative semidefiniteness in the selected frame for polarizations that satisfy the Coulomb and Lorentz conditions implies semidefiniteness in all reference frames with polarizations that satisfy the Coulomb and Lorentz conditions.

This completes the demonstration that the two terms given by (\ref{m-expnd}) have a decomposition (\ref{p1}), and that the constructed approximation to the Feynman series Compton scattering cross section satisfies the nonnegativity (\ref{4pt-trun}). The connected four-point function that results from $-i$ times the VEV derived from the Feynman rules with an approximation for $a_u$ is nonnegative. The nonnegativity applies in the subspace of photon polarizations satisfying the Lorentz and Coulomb conditions. A QFT that approximates the Feynman rules Compton scattering cross section results from a definition for the higher order connected functions. One organization supplements any $M$ and $B$ dependent factors in (\ref{trun-eval}) from Appendix \ref{app-vev} with factors that include connected four-point functions of the form (\ref{p1}).\[\renewcommand{\arraystretch}{1.25} \begin{array}{l} {^C\!W}_{n+m}(f_n^*\, g_m) := {\ds \frac{\overline{\varsigma_n}\, \varsigma_m}{n!m!}} \, c_{n+m}{\ds \sum_a \int} (d\xi)_{n+m}{\ds \int} du \; {\bf S}[ \overline{\tilde{f}_n}((\xi)_n)]\,{\bf S}[\tilde{g}_m((\xi)_{n+1,n+m})]\times\\
 \qquad \exp(i{\ds \sum_{\ell=1}^{n+m}}s_\ell p_\ell u)\, {\ds \prod_{k=1}^{n+m}} \delta^+_k \left(\frac{\ds \partial\;}{\ds \partial\rho_k} \right) \ldots \quad {\ds \prod_{b=1}^{N_b}}\left(1+{\ds \int} \mu_a(dv_b) \; \frac{\ds \partial\;}{\ds \partial\alpha_b} \frac{\ds \partial\;}{\ds \partial\alpha'_b} \right)\times\\
 \qquad \exp({\ds \sum_{\ell'=1}^{N_b}} \left(\alpha_{\ell'} {\ds \sum_{i<j}^n} \rho_i \rho_j \,\overline{T_a(v_{\ell'},\xi_i,\xi_j)}+\alpha'_{\ell'} {\ds \sum_{n<i<j}^{n+m}} \rho_i \rho_j \,T_a(v_{\ell'},\xi_i,\xi_j)\right))\end{array} \]with $N_b:=\left[\frac{n+m}{4}\right]$, the greatest integer less than or equal to $(n+m)/4$, and ${^C\!W}_{n+m}$ is evaluated for $(\rho)_{n+m}=(\alpha)_{N_b}=(\alpha')_{N_b}=0$.

% = = = = Error from approximation = = = =
The approximation improves for small momentum exchanges ${\bf p}_4-{\bf p}_2$. The fractional error in $|a_u|$ from (\ref{constrt-a}) contrasted with the first contributing order Feynman series value (\ref{feyn-a}) is\[\renewcommand{\arraystretch}{2.25}\begin{array}{rl} \left|{\ds \frac{ \frac{1}{(p_4\!-\!p_1)^2\!-\!m^2} -\frac{1}{(p_2\!-\!p_1)^2\!-\!m^2}}{\frac{1}{(p_4\!-\!p_1)^2\!-\!m^2}}}\right| &=\left|{\ds \frac{p_1(p_2 -p_4)}{p_1p_2}}\right|\\
 &={\ds \frac{\hat{\varrho}_1 (1-\cos \theta)}{m+2\hat{\varrho}_1}}\end{array}\]with $\theta$ the angle between the incident and product electron momenta in the center of momentum frame ($\varrho_2\varrho_4 \cos \theta := {\bf p}_2\cdot {\bf p}_4$ for the frame with ${\bf p}_1+{\bf p}_2=0$), and $\hat{\varrho}_1$ is the energy of the incident photon $p_1$ in the rest frame of the incident electron $p_2$. $\varrho_j^2:={\bf p}_j^2$. The fractional error is greatest when the product electrons are backscattered with respect to the incident electrons. $m\,\hat{\varrho}_1=\varrho_1(\varrho_1+\sqrt{m^2+\varrho_1^2})$ with $\varrho_1$ the incident photon energy in the center of momentum frame. 

%= = = = = = = = = = = = = = = = = = = = = = =
%= = = = = = = Appendices  = = = = = = = = = =
%= = = = = = = = = = = = = = = = = = = = = = =
\section{Appendices}
\renewcommand{\thesubsection}{\Alph{subsection}}
\setcounter{subsection}{0}

%= = = = = = = = = = = = = = = = = = = = = = =

\subsection{Notation and the VEV} \label{app-vev}

Spacetime coordinates are designated $x:=t,{\bf x}$, energy-momentum vectors are $p:=E,{\bf p}$ and more generally, Lorentz vectors are $q:=q_{(0)},{\bf q}$. $x,p,q\in {\bf R}^d$, ${\bf x},{\bf p},{\bf q}\in {\bf R}^{d-1}$, $x^2:=x^T g x=t^2-{\bf x}^2$, $px=p^T g x$ with $g$ the Minkowski signature matrix, ${\bf x}^2:=x_{(1)}^2\!+\!\ldots x_{(d-1)}^2$ is the square of the Euclidean length in ${\bf R}^{d-1}$, and\[E_j^2=\omega_j^2:=m_{\kappa_j}^2+{\bf p}_j^2\]describe mass shells. $p \in \bar{V}^+$, the closed forward cone, if $p^2 \geq 0$ and $E\geq 0$. In four dimensions, the components of ${\bf p}$ are alternatively designated as $p_{(1)},p_{(2)},p_{(3)}$ or $p_x,p_y,p_z$ as convenient. The components of Lorentz covariant vectors are designated as, for example, $p_{j;(k)}$ to distinguish Lorentz components $k$ from argument labels $j$. In a multiple argument function or generalized function $f(x_1, \ldots x_n)_{\kappa_1 \ldots \kappa_n}$, $m_{\kappa_j}$ is the mass associated with the $j$th argument and the $\kappa_j$th field component, $\kappa_j\in \{1,2,\ldots N_c\}$. Multiple arguments are denoted $(x)_n:=x_1,x_2\ldots x_n$ and $(x)_{k,n}:=x_k,\ldots x_n$ for either ascending or descending sequences of indices. The multiple argument notation includes recursion, for example,\[(\sum_{\nu}(\int d\zeta)_2)_3:= \sum_{\nu_1} \int d\zeta_1 \int d\zeta_2 \sum_{\nu_2} \int d\zeta_3 \int d\zeta_4 \sum_{\nu_3} \int d\zeta_5 \int d\zeta_6.\]Dirac delta generalized functions supported on mass shells are denoted\[\delta_j^{\pm}:= \frac{\delta(\pm E_j - \omega_j)}{2\omega_j},\qquad\qquad \hat{\delta}_j := \delta(p_j^2-m_{\kappa_j}^2) = \delta_j^+ + \delta_j^-.\]Sign conventions for Fourier transforms are set by the functions,\[ \tilde{f}_n((p)_n) := \frac{1}{(2 \pi)^{\frac{nd}{2}}}\,\int_{{\bf R}^{nd}} (dx)_n \; \prod_{k=1}^n e^{-ip_k x_k} f_n((x)_n),\]together with the definition of the Fourier transform of generalized functions $\tilde{T}(\tilde{f}) = T(f)$. Summation notation is used for generalized functions, $\int dx\; T(x) f(x):=T(f)$ for a generalized function $T(x)$ and an appropriate function $f(x)$. The summation\[\int_{-\infty}^{\infty} ds\; e^{-\alpha s^2 +\beta s} = \sqrt{\frac{\pi}{\alpha}} \; e^{\beta^2/(4 \alpha)}\]is used repeatedly. Notation includes the shorthand $(\xi)_n:=(p,\kappa)_n$, $(-\xi)_n:=(-p,\kappa)_n$ and\[\int (d\xi)_n:=\sum_{\kappa_1=1}^{N_c}\ldots \sum_{\kappa_n=1}^{N_c} \int dp_1\ldots dp_n.\]In particular, $\tilde{\Phi}_\ell:=\tilde{\Phi}(\xi_\ell)=\tilde{\Phi}(p_\ell)_{\kappa_\ell}$. $\overline{\alpha}$ denotes the complex conjugate of $\alpha \in {\bf C}$ and $\| A\|$ denotes the determinant of a square matrix $A$ or the Hilbert space norm of an element $A$, depending on context. $[M_{ij}]$ denotes the matrix with elements $M_{ij}\in {\bf C}$ or the $M_{ij}$ may be a $n$x$n$ set of $m$x$m$ matrices resulting in a $nm$x$nm$ matrix $[M_{ij}]$. The 2x2 Pauli spin matrices are designated\begin{equation}\label{paulispin} \renewcommand{\arraystretch}{1} \sigma_0:=\left( \begin{array}{cc} 1 & 0\\
 0 & 1\end{array} \right), \qquad
\sigma_1 := \left( \begin{array}{cc} 0 & 1\\
 1 & 0\end{array} \right), \qquad
\sigma_2 := \left( \begin{array}{cc} 0 & -i\\
 i & 0\end{array} \right), \qquad
\sigma_3 := \left( \begin{array}{cc} 1 & 0\\
 0 & -1\end{array} \right). \end{equation}

The units of the spacetime coordinates are length, and the units of the masses and energy-momentum coordinates are inverse length. Conversion of $t$ to units of time is then ``time'' = $t/c$. ``mass'' = $\hbar m/c$, ``momentum'' = $\hbar p$ and energies are, for example, ``energy'' = $\hbar c \sqrt{m^2+{\bf p}^2}$.

%= = = = = = = = = = VEV = = = = = 
The constructions [\ref{gej05},\ref{mp01}] are described by a Wightman-functional and a class of functions ${\cal B}\subset {\cal A}$. For every free field Wightman-functional $\underline{W_o}$, there is a family of nontrivial Wightman-functionals $\underline{W}$. One or more elementary particles are included in the description of the free field, $\underline{W_o}$. In the case of a free field, the only connected function that contributes is the two-point (generalized) function (\ref{twopoint}).

%= = = = the function class = = = = =
The Wightman-functional $\underline{W}$ is a sequence of local, Poincar\'{e} covariant generalized functions. $\underline{W}$ is a functional dual to terminating sequences of functions\[ \underline{f} := (\, f_0,\ldots, f_n((x)_n)_{\kappa_1\ldots \kappa_n} ,\ldots)\]with each $f_n((x)_n)_{\kappa_1\ldots \kappa_n}\in {\cal A}$ one of a sequence of $(N_c)^n$ $n$-argument functions. $f_0$ is a complex number. The algebra of function sequences ${\cal A}$ has the product (\ref{prod}), the $*$-map (\ref{dualf}) and is described in note \ref{foot-A} of the Introduction.
 
%= = = = higher order truncated (connected) functions = = = = =
The constructions include higher order connected functions.\begin{equation}\label{trun-eval} \renewcommand{\arraystretch}{1.25} \begin{array}{l} {^C\!W}_{n+m}(f_n^*\, g_m) := {\ds \frac{\overline{\varsigma_n}\, \varsigma_m}{n!m!}} \, c_{n+m}{\ds \int} (d\xi)_{n+m}{\ds \int} du \; \exp(i{\ds \sum_{\ell=1}^{n+m}}s_\ell p_\ell u)\, {\ds \prod_{k=1}^{n+m} \delta^+_k \left(\frac{\ds \partial\;}{\ds \partial\rho_k} \right)}\times\\
 \quad \exp({\ds \sum_{i<j}^n} \rho_i \rho_j \,\overline{U_n(p_i\!-\!p_j) M_{\kappa_i \kappa_j}(p_i\!-\!p_j)}+{\ds \sum_{n<i<j}^{n+m}} \rho_i \rho_j \,U_m(p_i\!-\!p_j) M_{\kappa_i \kappa_j}(p_i\!-\!p_j) ) \times\\
 \quad \exp ({\ds \sum_{i=1}^n \sum_{j=n+1}^{n+m}} \rho_i \rho_j \beta_{i+j-n}\Upsilon(p_i\!+\!p_j) (DB)_{\kappa_i \kappa_j}(p_{i}\!+\!p_j))\; {\bf S}[ \overline{\tilde{f}_n}((\xi)_n)]\,{\bf S}[\tilde{g}_m((\xi)_{n+1,n+m})]\end{array} \end{equation}evaluated at $(\rho)_{n+m}=0$ and with $s_\ell=-1$ for $\ell \leq n$ and $s_\ell=1$ otherwise. The signed symmetrization ${\bf S}[]$ in (\ref{permutation}) is described in Appendix \ref{app-trun}. The connected functions ${^C\!W}_n$ are identified as the connected contributions of the Wightman functions $W_n$.

The VEV of the construction [\ref{gej05}] are described by:\begin{list}{}{\itemsep -0.06in} \item[C1.] the free field two-point function (\ref{twopoint}) that determines the constituent elementary particles \item[C2.] coefficients $c_n$ that are the moments of a nonnegative measure \item[C3.] complex constants $\varsigma_n$ \item[C4.] Lorentz invariant functions $U_n(p),\Upsilon(p)$ that are multipliers of tempered functions \item[C5.] coefficients $\beta_j$ that are Laplace transforms (\ref{bmat}) of a nonnegative measure $\mu_{\beta}(dv)$ \item[C6.] a nonnegative, Lorentz invariant measure $\mu_s(ds)$ with\[B_{\kappa_k \kappa_j}(p) := \int \mu_s(ds)\; M_{\kappa_k \kappa_j}(s)\; e^{-sp};\]\item[C7.] a summation (\ref{permutation}) over signed permutations of arguments, ${\bf S}[]$.\end{list} Variations of these constructions include convex sums of the connected functions (\ref{trun-eval}) and additional organizations for Lorentz invariant functions and submatrices of $M(p)$ into nonnegative forms. Conditions (\ref{matcond}), (\ref{matlocal}), (\ref{matcond2}) and C1-C7, rather than equations of motion, describe the constructions.

%= = = = = = = = = = Hamiltonian = = = = =
The constructed VEV are Poincar\'{e} covariant solutions of the Klein-Gordon (Schr\"{o}dinger) equation and satisfy the spectral support condition. Consequently, evolution with time is analogous to the time evolution of a free field [\ref{gej05}].\begin{equation}\label{constr-hamil}\renewcommand{\arraystretch}{1.25} \begin{array}{l} \langle \underline{f}| U(t) \underline{g} \rangle = {\ds \sum_{n,m} \;\sum_{\kappa_1=1}^{N_c}\ldots \sum_{\kappa_{n+m}=1}^{N_c} \int} (dx)_{n+m}\; ((D\cdot)_n W_{n\!+\!m}((x)_{n\!+\!m}))_{\kappa_1\ldots \kappa_{n\!+\!m}}\times \\
 \qquad \qquad \qquad \qquad \overline{f_n}((x)_{n,1})_{\kappa_n \ldots \kappa_1} \; g_m((x_{(0)}-t,{\bf x})_{n+1,n+m})_{\kappa_{n+1} \ldots \kappa_{n+m}}\\
\qquad = {\ds \sum_{n,m} \int} d(\xi)_{n+m}\;((D\cdot)_n \tilde{W}_{n+m}((\xi)_{n+m})) \overline{\tilde{f}_n((-\xi)_{n,1})}{\ds \prod_{k=n+1}^{n+m}} e^{-i\omega_k t}\tilde{g}_m((\xi)_{n+1,n+m}) \end{array}\end{equation}for $\underline{f}, \underline{g} \in {\cal B}$. The interaction results from the mass shell singularities of the VEV. The free field Hamiltonian commutes with the angular momentum operators, generators of the Poincar\'{e} group, and angular momentum conservation follows from Poincar\'{e} covariance.

%= = = = = = = = = = = = = = = = = = = = = = =
\subsection{Functional-generators} \label{app-trun}

%= = = = = = = generator set up, definition of W = = = = =
The VEV described by (\ref{twopoint}) and (\ref{trun-eval}) are generalized function coefficients in a multinomial expansion that results from summation of symmetrized sums of products of two generator functionals. There is a generator ${\cal G}_o$ for the free field Wightman-functional $\underline{W_o}$ and a generator ${\cal G}_{n,m}$ for higher order connected functions. The Fourier transforms of the VEV are\begin{equation}\label{w-defn}\tilde{W}_n((\xi)_n):=\left({\ds \prod_{j=1}^n} \frac{\ds \partial\;}{\ds \partial\alpha_j}\right) \; {\ds \sum_{k=0}^n} \left( \frac{\ds {\bf S}[{\cal G}_{k,n-k}((\alpha,\xi)_n)\Theta_{k,n} {\cal G}_o((\alpha,\xi)_n)]}{\ds k!\, (n-k)!}\right)\end{equation}evaluated at $(\alpha)_n=0$. The energy ordering function $\Theta_{k,n}=1$ when $-E_j>0$ for every $j\leq k$ and $E_j>0$ for $k<j\leq n$ and $\Theta_{k,n}=0$ otherwise. This form exhibits a unique vacuum (is indecomposable [\ref{borchers}]). In [\ref{gej05}] it was demonstrated that $\tilde{W}_n((\xi)_n)=\tilde{W}_{o;n}((\xi)_n)$, the VEV of a free field, when the ${\cal G}_{n,m}=1$. It was also demonstrated that due to the limited energy supports, the signed permutations can be limited to within function arguments,\begin{equation}\label{H-G} \renewcommand{\arraystretch}{1.25} \begin{array}{rr} \underline{W}(\underline{f}^*\, {\bf x}\, \underline{g})&= 
{\ds \sum_{n,m} \frac{1}{n!m!} \int} (d\xi)_{n+m}\;{\bf S}[((D^T\cdot)_n \overline{\tilde{f}_n}((-\xi)_{n,1}))]\;{\bf S}[\tilde{g}_m((\xi)_{n+1,n+m})] \times\\
 & \left( {\ds \prod_{\ell=1}^{n+m}} {\ds \frac{\partial\;}{\partial \alpha_\ell}}\right)\, {\cal G}_{n,m}((\alpha,\xi)_{n\!+\!m})\, {\cal G}_o((\alpha,\xi)_{n\!+\!m}) \quad\end{array}\end{equation}when $\underline{f},\underline{g} \in {\cal B}$ and $(\alpha)_{n+m}=0$. The summation over signed permutations of arguments is denoted by\begin{equation}\label{permutation} {\bf S}[T_n((x)_n)_{\kappa_1 \ldots \kappa_n}] := \sum_{\pi} s_{\kappa_{\pi_1} \ldots \kappa_{\pi_n}} \,T_n(x_{\pi_1},x_{\pi_2},\ldots x_{\pi_n})_{\kappa_{\pi_1} \ldots \kappa_{\pi_n}}.\end{equation}The summation includes all $n!$ permutations of $1$ through $n$. The signs $s_{\kappa_{\pi_1} \ldots \kappa_{\pi_n}}$ are determined by transpositions, $s_{\ldots \kappa_j \kappa_{j+1}\ldots }= \sigma_{\kappa_j \kappa_{j+1}} s_{\ldots \kappa_{j+1} \kappa_j\ldots }$ with $\sigma_{\kappa_j \kappa_{j+1}}=- 1$ if $\kappa_j, \kappa_{j+1} > N_b$, and $\sigma_{\kappa_j \kappa_{j+1}}= 1$ otherwise. $s_{\ldots \kappa_j \kappa_{j+1}\ldots }=0$ when $\kappa_j=\kappa_{j+1}>N_b$ and $|s_{\kappa_{\pi_1} \ldots \kappa_{\pi_n}}|=1$ when no $\kappa_i=\kappa_j>N_b$ for $i\neq j$. These signs agree with the commutation relations of the free field that apply when $x_i-x_j$ is space-like. The argument of ${\bf S}[\cdot]$ indicates a term with positive sign and, together with the transpositions, determines the signs $s_{\kappa_{\pi_1} \ldots \kappa_{\pi_n}}$.

% = = = = Generators = = = = = = = = = =
The generator for the free field is constructed as the multinomial in $(\alpha)_{n+m}$ with generalized function coefficients that results in\begin{equation}\label{wodefn-free} \renewcommand{\arraystretch}{1.25} \begin{array}{l} {\ds \prod_{j\in I_{\ell,n}}} \frac{\ds \partial\;}{\ds \partial\alpha_j}\, {\cal G}_o((\alpha,\xi)_n):=\tilde{W}_{o;\ell}((\xi)_{I_{\ell,n}},(\xi)_{I_{\ell',m}})\\
 \qquad \qquad := \left\{ \renewcommand{\arraystretch}{1.25} \begin{array}{ll} {\ds \sum_{\mathit{pairs}}} s_{\pi_1\ldots \pi_{2k}} \tilde{\Delta}(\xi_{\pi_1},\xi_{\pi_2})\ldots \tilde{\Delta}(\xi_{\pi_{2k\!-\!1}},\xi_{\pi_{2k}}) & \quad \ell=2k\\
 0 & \quad \ell=2k\!+\!1 \end{array} \right. \end{array}\end{equation}when $(\alpha)_n=0$. $s_{\pi_1\ldots \pi_{2k}}$ is from (\ref{permutation}) with $s_{12\ldots n}=1$ and the sum is over all $(2k)!/(2^k k!)$ pairs from $I_{2k,n}$ without regard to order. The $\ell$ elements of the set of integers $I_{\ell,n}:=\{i_1,i_2,\ldots i_{\ell}\}$ are distinct elements of $\{1,2,\ldots n\}$. The indices of the two-point functionals are in ascending index order, $\pi_j<\pi_k$ when $j<k$.
 
The higher order connected functions result from ${\cal G}_{n,m}((\alpha,\xi)_{n\!+\!m})$ that are Hadamard functions of the $M(p)_{\kappa_1 \kappa_2}$ from (\ref{twopoint}).\begin{equation}\label{genr2} \renewcommand{\arraystretch}{1.25} \begin{array}{l} \ln\left({\cal G}_{n,m}((\alpha,\xi)_{n+m})\right) := {\ds \int} d\zeta_1 \; \overline{z_n((-\xi)_{n,1},M^*)}\,z_m((\xi)_{n\!+\!1,n\!+\!m},M)\times\\
 \qquad \qquad \exp({\ds \int} d\zeta_2 \; \overline{w_n((-\xi)_{n,1},DC)}\,w_m((\xi)_{n\!+\!1,n\!+\!m},C)) \end{array}\end{equation}with $DM=C^*C$ from (\ref{matcond}) and\begin{equation}\label{z-defn2} \renewcommand{\arraystretch}{1.25} \begin{array}{rl} z_n((\xi)_{\eta\!+\!1,\eta\!+\!n},M) :=& \varsigma_n {\ds \prod_{\ell=\eta+1}^{\eta+n}} (a_\ell+ \lambda \alpha_{\ell} e^{-ip_{\ell} u } \delta^+_{\ell} \frac{\ds \partial\;}{\ds \partial\rho_{\ell}})\times\\
 &\qquad \qquad \exp({\ds \sum_{\eta<k<j}^{\eta+n}} \rho_k \rho_j \,U_n(p_k\!-\!p_j) M_{\kappa_k \kappa_j}(p_k\!-\!p_j) )\\
% \qquad =&z_n((\xi)_n,M;\zeta_1,(\alpha)_n) \\
 w_n((\xi)_{\eta\!+\!1,\eta\!+\!n},C):=& {\ds \sum_{j=\eta+1}^{\eta+n}} \;e^{-(j-\eta)v\!-\!p_j(s'\!+\!s)} \rho_j C(s)_{\ell \kappa_{j}}. \end{array}\end{equation}The generator is evaluated at $(\rho)_{n+m}=0$ and ${\cal G}_{n,m}=1$ for $n,m=0,1$. The parameters of $z_n$ include the $(\alpha)_{\eta\!+\!1,\eta\!+\!n}$ and $\zeta_1:=\lambda,u$, and for $w_n$ the parameters include $\zeta_2:=s',s,v,\ell$. The $\varsigma_n$ are complex constants. The real constants\[a_\ell =\left\{\begin{array}{ll}0\qquad&\ell=1,2\\1 &\mbox{otherwise}\end{array}\right.\]result in a lowest contributing term in the generator ${\cal G}_{n,m}$ that is quartic in the $(\alpha)_{n+m}$ and removes the divergent two-point contribution that would result from extrapolating (\ref{genr2}) to quadratic terms. $(\xi)_{n,1}$ indicates that the indices are in descending order. The indicated summations are\[\renewcommand{\arraystretch}{1.25} \begin{array}{l} {\ds \int d\zeta_1 := \int \sigma(d\lambda)\int du}\\ {\ds \int d\zeta_2 := \int \mu_u(ds') \int \mu_s(ds)\int \mu_\beta(dv) \sum_{\ell=1}^{N_c}}. \end{array}\]$\mu_s(ds)$ and $\mu_u(ds)$ are nonnegative, Lorentz invariant measures with support only for positive energies. These measures correspond with one-dimensional nonnegative tempered measures $\mu_1(d\lambda)$ [\ref{steinmann}] as\[ \mu_s(ds)=\left(a\delta(s)+\int \mu_1(d\lambda)\; \delta^+(s^2-\lambda)\right)\; ds.\]$\sigma(d\lambda)$ is a nonnegative measure with finite moments,\[c_n:=\int \sigma(d\lambda)\; \lambda^n.\]Also\begin{equation}\label{bmat} \renewcommand{\arraystretch}{1.25} \begin{array}{rl}B_{\kappa_k \kappa_j}(p) :=&{\ds \int} \mu_s(ds)\; M_{\kappa_k \kappa_j}(s)\; e^{-sp}\\ \Upsilon(p) :=&{\ds \int} \mu_u(ds)\; e^{-sp}\\
\beta_j:=&{\ds \int} \mu_\beta(dv)\; e^{-jv}.\end{array}\end{equation}The functions $\Upsilon(p)$ and $U_n(p)$ in (\ref{z-defn2}) are Lorentz invariants and multipliers of test functions. The $U_n(p),\Upsilon(p)$ attributed to each constituent matrix in $M(p)$ that are direct sum compositions may be distinct. In particular, the $U_n(p),\Upsilon(p)$ that apply for fermions and bosons may be distinct due to (\ref{matcond}).

%= = = = = = = = = = = = = = = = = = = = = = =
\subsection{Definitions for electrodynamics} \label{app-electro}

Electrodynamics includes a neutral, Lorentz vector field $A(x)=A(x)^*$ coupled with the two bispinor fields $\Psi(x)$ and $\Psi(x)^*$ of a charged fermion field. The two-point function (\ref{twopoint}) is determined to replicate the Feynman propagators from quantum electrodynamics (QED). The constructed VEV are composed from a two-point function described using the matrices $M(p)$ and Dirac adjoint matrix $D$ that satisfy conditions (\ref{matcond}), (\ref{matlocal}), and (\ref{matcond2}). The field has constituents,\begin{equation}\label{electro-field}\Phi(x):=\left(\begin{array}{c} A(x) \\ \Psi(x)\\ \Psi(x)^*
\end{array}\right)\end{equation}with $A(x),\Psi(x),\Psi^*(x)$ each four component fields when specializing to 3+1 spacetime, $d=4$.

The two-point VEV derives from a 12x12 matrix of generalized functions, $M(p)$. For electrodynamics, from (\ref{matcond}),\[M(p) = \left( \begin{array}{cc} M_1(p) &0\\
0 & M_2(p)\end{array}\right)\]with $M_1(p)$ the 4x4 photon component and $M_2(p)$ the 8x8 electron-positron component. The Dirac conjugation matrix is\begin{equation}\label{dconj} \renewcommand{\arraystretch}{1} 
D := \left( \begin{array}{cc} 1 & 0\\
 0 & D_2\\\end{array} \right)\end{equation}with a 4x4 identity matrix $1$, the 8x8 $D_2$ is defined below and for electrodynamics, $D=\overline{D}$.

The representations of the Poincar\'{e} group include interpretation of the states as particle species and polarizations. $\tilde{f}(p)_\kappa= w(p)_{\kappa} \tilde{f}(p)$ for $\kappa\in \{1,N_c\}$ with $\tilde{f}(p)\in {\cal B}$ and multipliers $w(p)_{\kappa}$ provides a convenient description of particle species and polarizations decoupled from the plane wave limit. For electrodynamics,\begin{equation}\label{polarzn} w(p) := \left( \begin{array}{c} w_\epsilon(p)\\ 0\\ 0\end{array}\right) \qquad \qquad w(p) := \left( \begin{array}{c} 0 \\ 0\\ w_p(p) \end{array}\right) \qquad \qquad w(p) := \left( \begin{array}{c} 0 \\ w_a(p)\\ 0 \end{array}\right) \end{equation}for photons $\epsilon$, electrons $p$, and positrons $a$ respectively. The notation distinguishes argument labels from the array components as $w_{\alpha,j}(p)_{\kappa}$. $\kappa$ labels elements within a four element polarization array for argument $j$ and particle type $\alpha=\epsilon,a,p$. The segregation of elements associated with the bosons and fermions is stable under Lorentz transformation since the Lorentz transformations reduce into 4x4 constituents as\begin{equation}\label{loren-electro}S(A) = \left( \begin{array}{ccc} S_1(A) &0&0\\
0 & \overline{S_p}(A)&0\\
0&0&S_p(A)\end{array}\right).\end{equation}

A free field development establishes correspondence definitions for field components with particle species and polarizations. In the case of free fields, the semi-norm (\ref{norm}) applies in ${\cal A}$, the $*$-involution is an automorphism of ${\cal A}$, and as a consequence the free field is a Hermitian Hilbert space operator [\ref{cook}].

%= = = = = = = = = =
\subsubsection{Polarization of photon states and the free field}

%= = = = = = = = = = = =
% Boson M, D, S defns
%= = = = = = = = = = = =
An established development of electrodynamics [\ref{weinberg},\ref{schwabl}] quantizes the vector potential and relies on local gauge invariance to achieve a covariant development. The photon propagator in the Feynman series results from the two-point function (\ref{twopoint}) with\begin{equation}\label{m1-form}M_1(p)=-2\pi\, g,\end{equation}and $g:=\mbox{diag}(1,-1,-1,-1)$, the Minkowski signature. The Lorentz transformations are\begin{equation}\label{lorentz1} S_1(A) = \Lambda(A)^{-1}\end{equation}with $A\in$SL(2,C), the group of 2x2 complex matrices of determinant one and covering group of the proper orthochronous Lorentz group. The four dimensional Lorentz transformation associated with $A$ is $\Lambda(A)_{\mu\nu}=\frac{1}{2} \mbox{Trace}(\sigma_\mu A\, \sigma_\nu A^*)$ using the Pauli spin matrices (\ref{paulispin}).

%= = polarization description = = =
Difficulties with this development include that $-g$ is indefinite and there are only two, transversely polarized states of photons observed. $-g$ is positive semidefinite in the subspace of photon polarization states that satisfy the Coulomb condition, $w_\epsilon(p)_0=0$. Limiting states of the photons to two transverse polarizations is not evidently Lorentz covariant and covariance relies on an additional property, local gauge invariance of Maxwell's equations. This development does not refer to equations of motion and it is taken as a constraint that all photon states satisfy the Coulomb and Lorentz conditions. The covariant Lorentz condition is that $w_\epsilon(p)\,p=0$ with $p$ the energy-momentum of a plane wave photon. For electrodynamics, the polarization of photon states is restricted to those with a semi-norm. This restriction is similar to the limitation on functions in ${\cal B}$ to positive energy support to achieve a semi-norm except this polarization constraint is not Lorentz invariant.

For the free photon field, $a_r(q)$ designates the photon annihilation operator, and $a_r^*(q)$ designates the photon creation operator. $a_r^*(q)$ creates a boson in a plane wave state with momentum ${\bf q}$, energy $\sqrt{{\bf q}^2}$, and a polarization $w_\epsilon(q)$ described below. The free fields,\begin{equation}\label{photon-field}A(x) =A^+(x)+A^-(x),\end{equation}are solutions to the Klein-Gordon (Schr\"{o}dinger) equation with Fourier transforms\begin{equation}\label{photon-f}\renewcommand{\arraystretch}{1.25} \begin{array}{ll} \tilde{A}^+(-p)_\kappa &= \sqrt{2\pi}\, \delta(E-\omega) {\ds \sum_r}\, \epsilon_r(p)_{\kappa} a_r(p)\\
 \tilde{A}^-(p)_\kappa &= \sqrt{2\pi}\, \delta(E-\omega) {\ds \sum_r}\, \overline{\epsilon}_r(p)_{\kappa} a^*_r(p)\end{array}\end{equation}and\[(A^+(x)_\kappa)^* = A^-(x)_\kappa.\]Then,\[ A(x)_\kappa = {\ds \sum_r \int}{\ds \frac{d{\bf p}}{(2\pi)^{3/2}}} \;\left( e^{i\omega t-i{\bf p}\cdot {\bf x}} \,\epsilon_r(p)_{\kappa} a_r(p)+ e^{-i\omega t+i{\bf p}\cdot {\bf x}} \,\overline{\epsilon}_r(p)_{\kappa} a^*_r(p)\right),\]and in the Gupta-Bleuler development [\ref{schwabl}], the commutation relations are\[\renewcommand{\arraystretch}{1.25} \begin{array}{rl} [a_r(q),a_{r'}^*(q')] &= -g_{rr'}\,\delta({\bf q}-{\bf q}')\\
 \left[ a_r(q),a_{r'}(q') \right] &=0.\end{array}\]These Gupta-Bleuler commutation relations differ by a sign from canonical commutation relations.

Evaluation of the two-point function (\ref{twopoint}) results in a quadratic expression for the free photon field expansion coefficients in terms of $M_1(p_2)$.\begin{equation}\label{2pt-photon}\renewcommand{\arraystretch}{1.25} \begin{array}{rl} {\ds \frac{1}{2\pi}}\,M_1(p_2)_{\kappa_1\kappa_2} &=2\omega_2\,{\ds \sum_r}(-g_{rr})\, \epsilon_r(p_2)_{\kappa_1} \overline{\epsilon}_r(p_2)_{\kappa_2}\\
 &=-g_{\kappa_1\kappa_2}\end{array}\end{equation}from (\ref{m1-form}). One basis for the free photon field expansion coefficients has\begin{equation}\label{photon-pol1} \epsilon_0(p) =\frac{1}{\sqrt{2\omega}} \,\left(\begin{array}{c}1 \\ 0\\ 0\\ 0\end{array}\right),
\qquad \epsilon_{3}(p) =\frac{1}{\sqrt{2\omega}} \,\left(\begin{array}{c}0 \\ u_x\\ u_y\\ u_z\end{array}\right)\end{equation}
with ${\bf u}={\bf p}/\sqrt{{\bf p}^2}=(u_x,u_y,u_z)$, the unit vector in the direction of the momentum. One set of basis vectors that span the subspace of transverse polarizations is\begin{equation}\label{photon-pol2} \epsilon_1(p) =\frac{1}{\sqrt{2\omega}} \,\left(\begin{array}{c}0\\ u_{yz}\\ -u_xu_y/u_{yz}\\ -u_xu_z/u_{yz}\end{array}\right), \qquad \epsilon_2(p) =\frac{1}{\sqrt{2\omega}} \,\left(\begin{array}{c}0\\ 0\\ u_z/u_{yz}\\ -u_y/u_{yz}\end{array}\right) \end{equation}with $u_{yz}\geq 0$ and $u_{yz}^2:=u_y^2+u_z^2$. The photon field expansion coefficients (\ref{photon-pol1}) and (\ref{photon-pol2}) are orthogonal,\[2\omega \sum_\kappa \overline{\epsilon_r}(p)_{\kappa} \epsilon_{r'}(p)_{\kappa} =\delta_{r,r'}.\]

Using (\ref{photon-field}) and (\ref{photon-f}), the creation and annihilation operators are expressed in terms of the field. For the creation operator,\[2\omega i \sum_r \overline{\epsilon_r}(q)_{\kappa} a_r^*(q) =\int \frac{d{\bf x}}{(2\pi)^{3/2}}\; e^{iqx}(iq_{(0)}A(x)_\kappa-\dot{A}(x)_\kappa)\]with $q$ on the positive energy mass shell and dot represents the time derivative. The orthogonality of the photon field expansion coefficients is used to solve for the creation operator.\begin{equation}\label{photon-c}a_r^*(q)=\sum_\kappa \int dp\; \tilde{A}(p)_\kappa\tilde{\ell}^c(0;p)_\kappa \end{equation}for a plane wave limit of LSZ functions\[\renewcommand{\arraystretch}{1.75} \begin{array}{rl} \tilde{\ell}^c(0;p)_\kappa&={\ds \frac{(\omega+E)}{\sqrt{2\pi}}}\, e^{i(\omega-E)t}\epsilon_r(p)_{\kappa} \delta({\bf p}-{\bf q})\\
 &={\ds \frac{2\omega}{\sqrt{2\pi}}}\, \epsilon_r(p)_{\kappa} \delta({\bf p}-{\bf q})\end{array}\]since $p$ is on the positive mass shell. These $\ell(x)_\kappa \in {\cal B}$ and the dependence on ${\bf q}$ and $r$ is suppressed in the notation. The polarization $w_\epsilon(q)$ of the state created by $a_r^*(q)$ is then $\epsilon_r(q)$. A similar development results in the annihilation operator.\[a_r(q) =\sum_\kappa \int dp\; \tilde{A}(-p)_\kappa\tilde{\ell}^a(0;p)_\kappa\]for the LSZ function\[\tilde{\ell}^a(0;p)_\kappa ={\ds \frac{2\omega}{\sqrt{2\pi}}}\, \overline{\epsilon_r}(p)_{\kappa} \delta({\bf p}-{\bf q})\]since $p$ is on the mass shell.
 
The transversely polarized states that satisfy (\ref{pold}) are\begin{equation}\label{pol-st-photon} w_{\epsilon}(p) = a_{1}\epsilon_1(p)+a_{2}\epsilon_2(p)\end{equation}with $|a_{1}|^2+|a_{2}|^2=1$. Only the observed, transversely polarized states are considered. For photon states that satisfy the Coulomb and Lorentz conditions, $\overline{w}_\epsilon(p_k)^T DM(p_k) w_\epsilon(p_k)=\overline{w_{\epsilon}}(p_k)^T w_{\epsilon}(p_k)\geq 0$.

%= = = = = = = = = =

\subsubsection{Polarization of electron states and the free field}

%= = = = = = = = = = = =
% Fermion M, D, S defns
%= = = = = = = = = = = =
Electrodynamics [\ref{weinberg},\ref{schwabl}] includes a charged, spin-1/2 free field realized as two bispinor fields. The propagators used in Feynman series for electrodynamics result from the two-point function (\ref{twopoint}) with\begin{equation}\label{M2-defn}\renewcommand{\arraystretch}{2.75} \begin{array}{l} M_2(p)= 2\pi\, \left({\renewcommand{\arraystretch}{1}\begin{array}{cc}
0 & ({\not \! p}+m)\gamma_0\\
\gamma_0({\not \! p}-m)^T & 0 \end{array}}\right), \qquad D_2=\left({\renewcommand{\arraystretch}{1}\begin{array}{cc} 0 & 1\\
1 & 0\end{array}} \right)\\
D_2 M_2(p)= 2\pi\, \left({\renewcommand{\arraystretch}{1}\begin{array}{cc}
\gamma_0({\not \! p}-m)^T & 0\\
0 & ({\not \! p}+m)\gamma_0 \end{array}}\right)\end{array}\end{equation}expressed in 4x4 components for a field with eight components, $\Phi=(\Psi, \Psi^*)$ from (\ref{electro-field}). ${\not \! p}$ is the 4x4 matrix\begin{equation}\label{slash} \renewcommand{\arraystretch}{2} \begin{array}{rl} {\not \! p} &:={\ds \sum_{k=0}^3} p_{(k)}\, \gamma_k^*\\
 &= \left( \renewcommand{\arraystretch}{1} \begin{array}{cc} E\sigma_0 & -P(0,{\bf p})\\
 P(0,{\bf p}) & -E\sigma_0\end{array} \right)\end{array}\end{equation}with gamma matrices represented\begin{equation}\label{gamma} \gamma_0=\left(\begin{array}{cc} \sigma_0 & 0\\ 0 &-\sigma_0\end{array}\right), \qquad \qquad \gamma_j=\left(\begin{array}{cc} 0 & \sigma_j\\ -\sigma_j & 0\end{array}\right)\end{equation}for $j=1,2,3$ and\begin{equation}\label{P-fermion} P(p) := {\ds \sum_{k=0}^3} p_{(k)}\, \sigma_k =\left( \renewcommand{\arraystretch}{1} \begin{array}{cc} E\!+\!p_{(3)} & p_{(1)}\!-\!ip_{(2)} \\ p_{(1)}\!+\!ip_{(2)} & E\!-\!p_{(3)}\end{array} \right),\end{equation}using the Pauli spin matrices $\sigma_k$ from (\ref{paulispin}). The $\sigma_k$ and $P(p)$ are Hermitian 2x2 matrices. From (\ref{gamma}), \begin{equation}\label{gamma-star}\gamma_\mu^* = g_{\mu \mu} \gamma_\mu=\gamma_0\gamma_\mu \gamma_0=\gamma_\mu^{-1}.\end{equation}When $p^2\geq m^2$, $D_2M_2(p)$ is a positive semidefinite matrix.

The Lorentz transformations $S_2(A)M_2(p)S_2(A)^T=M_2(\Lambda^{-1}p)$ in (\ref{loren-electro}) are\begin{equation}\label{lorentz-fermion-full}S_2(A) =\left( \renewcommand{\arraystretch}{1.25} \begin{array}{cc} \overline{S_p}(A) & 0\\
0 & S_p(A)\end{array} \right)\end{equation}with\begin{equation}\label{lorentz-fermion} \renewcommand{\arraystretch}{1.25} \begin{array}{rl} \overline{S_p}(A) &:={\ds \frac{1}{2}} \left( \renewcommand{\arraystretch}{1.25} \begin{array}{cc} A^{-1}+A^* & A^{-1}-A^*\\
A^{-1}-A^* & A^{-1}+A^*\end{array} \right)\end{array}\end{equation}and $A\in$SL(2,C) with $AP(p)A^*=P(\Lambda(A) p)$. This representation of the Lorentz group, selected for contrasts with Feynman rules results, is developed in Appendix \ref{app-eqrep}.
%= = = = = = = =

For the spin one-half fermions, $b_r(q),d_r(q)$ designate the free field annihilation operators, and $b_r^*(q),d_r^*(q)$ designate the free field creation operators. $b_r^*(q)$ creates an electron in a plane wave state with momentum ${\bf q}$, energy $\sqrt{m^2+{\bf q}^2}$, and a polarization bispinor $w_{p,r}(q)$ described below. Spins are linear superpositions of spin up and spin down and $r=1,2$. $d_r^*(q)$ creates a positron. The free fields,\begin{equation}\label{dirac-field}\Psi(x) =\Psi^+(x)+\Psi^-(x),\qquad \Psi(x)^* =\Psi^+(x)^*+\Psi^-(x)^*\end{equation}are solutions of the Dirac equations with Fourier transforms\begin{equation}\label{dirac-f}\renewcommand{\arraystretch}{1.25} \begin{array}{ll} \tilde{\Psi}^+(-p)_\kappa &= \sqrt{2\pi}\, \delta(E-\omega) \, (u_1(p)_{\kappa} b_1(p)+u_2(p)_{\kappa} b_2(p))\\
 \tilde{\Psi}^-(p)_\kappa &= \sqrt{2\pi}\, \delta(E-\omega) \, (v_1(p)_{\kappa} d^*_1(p)+v_2(p)_{\kappa} d^*_2(p)).\end{array}\end{equation}and\begin{equation}\label{dirac-fadj}\renewcommand{\arraystretch}{1.25} \begin{array}{ll} \tilde{\Psi}^+(p)^*_\kappa &= \sqrt{2\pi}\, \delta(E-\omega) \,(\overline{u}_1(p)_{\kappa} b^*_1(p)+\overline{u}_2(p)_{\kappa} b^*_2(p))\\
 \tilde{\Psi}^-(-p)^*_\kappa &= \sqrt{2\pi}\, \delta(E-\omega) \,(\overline{v}_1(p)_{\kappa} d_1(p)+\overline{v}_2(p)_{\kappa} d_2(p)).\end{array}\end{equation}Then,\[\renewcommand{\arraystretch}{1.25} \begin{array}{ll} \Psi(x)_\kappa &= {\ds \sum_{r=1}^2 \int}{\ds \frac{d{\bf p}}{(2\pi)^{3/2}}} \; \left( e^{i\omega t-i{\bf p}\cdot {\bf x}}\, u_r(p)_{\kappa} b_r(p)+ e^{-i\omega t+i{\bf p}\cdot {\bf x}}\, v_r(p)_{\kappa} d^*_r(p)\right)\\
 \Psi(x)^*_\kappa &= {\ds \sum_{r=1}^2 \int}{\ds \frac{d{\bf p}}{(2\pi)^{3/2}}} \; \left( e^{-i\omega t+i{\bf p}\cdot {\bf x}}\, \overline{u}_r(p)_{\kappa} b^*_r(p)+ e^{i\omega t-i{\bf p}\cdot {\bf x}}\, \overline{v}_r(p)_{\kappa} d_r(p)\right),\end{array}\]and the anticommutation relations are\[\renewcommand{\arraystretch}{1.25} \begin{array}{rl} [b_r(q),b_{r'}^*(q')]_+ &= [d_r(q),d_{r'}^*(q')]_+ =\delta_{rr'} \delta({\bf q}-{\bf q}')\\
 \left[b_r(q),b_{r'}(q')\right]_+ &= [d_r(q),d_{r'}(q')]_+= [d_r(q),b_{r'}(q')]_+= [d_r(q),b_{r'}^*(q')]_+=0.\end{array}\]

Evaluation of the two-point function (\ref{twopoint}) results in quadratic expressions for the free fermion field expansion coefficients in terms of $M_2(p_2)$.\begin{equation}\label{2pt-PP*}\renewcommand{\arraystretch}{1.75} \begin{array}{rl}\langle \Omega | \Psi(x_1)_{\kappa_1} \Psi(x_2)^*_{\kappa_2} \Omega\rangle &= {\ds \int} {\ds \frac{d{\bf p}_2}{(2\pi)^3}} \; e^{i\omega_2(t_1-t_2)-i{\bf p}_2\cdot ({\bf x}_1-{\bf x}_2)}\; {\ds \sum_r} u_r(p_2)_{\kappa_1} \overline{u}_r(p_2)_{\kappa_2}\\
  (({\not \! p}_2+m)\gamma_0)_{\kappa_1\kappa_2}&= 2\omega_2\,{\ds \sum_r} u_r(p_2)_{\kappa_1} \overline{u}_r(p_2)_{\kappa_2}\end{array}\end{equation}and\begin{equation}\label{2pt-P*P}\renewcommand{\arraystretch}{1.75} \begin{array}{rl}\langle \Omega | \Psi(x_1)^*_{\kappa_1}\Psi(x_2)_{\kappa_2} \Omega\rangle &= {\ds \int} {\ds \frac{d{\bf p}_2}{(2\pi)^3}} \; e^{i\omega_2(t_1-t_2)-i{\bf p}_2\cdot ({\bf x}_1-{\bf x}_2)} \; {\ds \sum_r} \overline{v}_r(p_2)_{\kappa_1} v_r(p_2)_{\kappa_2}\\
 (\gamma_0({\not \! p}_2-m)^T)_{\kappa_1\kappa_2}&= 2\omega_2\,{\ds \sum_r} \overline{v}_r(p_2)_{\kappa_1} v_r(p_2)_{\kappa_2}. \end{array}\end{equation}There are no contributions from\[\langle \Omega | \Psi(x_1)_{\kappa_1}\Psi(x_2)_{\kappa_2} \Omega\rangle =\langle \Omega | \Psi(x_1)^*_{\kappa_1}\Psi(x_2)^*_{\kappa_2} \Omega\rangle =0.\]

The fermion free field expansion coefficients $u_r(p),v_r(p)$ used in (\ref{dirac-f}) and (\ref{dirac-fadj}) select representations from the reducible representation of the Lorentz group (\ref{lorentz-fermion-full}) and are developed from descriptions of polarization and type, electron or positron, in the rest frames of the finite mass fermions. In the rest frame, with $p=(m,0,0,0)$ designated below as $p=0$, a realization of the spin states is\[u_1(0) =\left(\begin{array}{r}1\\ 0\\ 0\\ 0\end{array}\right),
\quad u_2(0) =\left(\begin{array}{c}0\\ 1\\ 0\\ 0\end{array}\right), \quad v_1(0) =\left(\begin{array}{r} 0\\ 0\\ 0\\ -1\end{array}\right),
\quad v_2(0) =\left(\begin{array}{r}0\\ 0\\ 1\\ 0\end{array}\right).\]$r=1$ is a spin up electron or positron respectively, and $r=2$ is spin down. From (\ref{matcond2}), (\ref{M2-defn}), (\ref{2pt-PP*}) and (\ref{2pt-P*P}),\begin{equation}\label{DM(0)} \renewcommand{\arraystretch}{1} \begin{array}{rl} {\ds \frac{1}{2\pi}}\, D_2M_2(p)&={\ds \frac{1}{2\pi}}\, \overline{S}_2(A)D_2M_2(0)S_2(A)^T\\
 &=2m\,{\ds \sum_{r=1}^2}\; \left(\renewcommand{\arraystretch}{1} \begin{array}{cc} S_p & 0\\
0 & \overline{S}_p \end{array}\right) \left(\begin{array}{cc} \overline{v_r}(0)v_r(0)^T &0\\ 0 & u_r(0)\overline{u_r}(0)^T\end{array}\right)\left(\begin{array}{cc}
S_p^* & 0\\
0 & S_p^T\end{array}\right)\\
 &= 2\omega\,{\ds \sum_{r=1}^2}\; \left(\begin{array}{cc} \overline{v_r}(p)v_r(p)^T &0\\ 0 & u_r(p)\overline{u_r}(p)^T\end{array}\right).\end{array}\end{equation}From (\ref{lorentz-fermion-full}), (\ref{2pt-PP*}) and (\ref{2pt-P*P}) are satisfied for general $p$ when\begin{equation}\label{ferm-pol}v_r(p) ={\ds \sqrt{\frac{m}{\omega}}}\; \overline{S}_p(A) v_r(0), \qquad u_r(p) ={\ds \sqrt{\frac{m}{\omega}}}\; \overline{S}_p(A) u_r(0)\end{equation}with $A$ a Lorentz transformation that takes $(m,0,0,0)$ to $p$, that is $(m,0,0,0)=\Lambda^{-1}p$ and $AP(m,0)A^*=mAA^*=P(\Lambda(m,0))=P(p)$. The pure boost is\[A = {\ds \frac{1}{\sqrt{2m(E+m)}}}\;(P(p)+m\sigma_0).\]

Polarization states (\ref{polarzn}) that are normalized linear combinations of Lorentz transforms of $u_r(0),v_r(0)$ are orthogonal to $u_r(p),v_r(p)$. This orthogonality enables inversion of (\ref{dirac-f}) and (\ref{dirac-fadj}) for the creation and annihilation operators in terms of the fields and identifies the polarization of states. A basis of polarization states $w_{p,r}(p),w_{a,r}(p)$ are distinguished as\begin{equation}\label{pol-st-fermion}\renewcommand{\arraystretch}{2.25}\begin{array}{rl}w_{p,r}(p)&:={\ds \frac{1}{2\sqrt{\omega m}}} \,S_p^T(A)^{-1} u_r(0)\\
w_{a,r}(p)&:={\ds \frac{1}{2\sqrt{\omega m}}} \, S^*_p(A)^{-1} v_r(0)\end{array}\end{equation}for $r=1,2$. $r=1$ is spin up and $r=2$ spin down for the rest frame z-axis, $u$ indicates an electron and $v$ indicates a positron. Then using (\ref{ferm-pol}),\begin{equation}\label{pol-basis}\sum_{k=0}^3 \overline{u_r}(p)_k w_{p,r'}(p)_k=\sum_{k=0}^3 v_r(p)_k w_{a,r'}(p)_k=\frac{\delta_{r,r'}}{2\omega}.\end{equation}From (\ref{dirac-field}), (\ref{dirac-f}) and (\ref{dirac-fadj}), the creation and annihilation operators can be expressed in terms of the free fields. For the electron creation operator,\[2\omega i \sum_r \overline{u_r}(q)_{\kappa} b_r^*(q) =\int \frac{d{\bf x}}{(2\pi)^{3/2}}\; e^{iqx}(iq_{(0)}\Psi(x)^*_\kappa-\dot{\Psi}(x)^*_\kappa)\]with $q$ on the positive energy mass shell and dot representing the time derivative. The orthogonality of the field expansion coefficients and polarization states (\ref{pol-basis}) is used to solve for the creation operator.\begin{equation}\label{electron-c} b_r^*(q) =\sum_\kappa \int dp\; \tilde{\Psi}(p)^*_\kappa \tilde{\ell}^c(0;p)_\kappa\end{equation}for a function that is the plane wave limit of LSZ functions.\[\renewcommand{\arraystretch}{1.75} \begin{array}{rl} \tilde{\ell}^c(0;p)_\kappa&={\ds \frac{(\omega+E)}{\sqrt{2\pi}}}\, e^{i(\omega-E)t}w_{p,r}(p)_{\kappa} \delta({\bf p}-{\bf q})\\
 &={\ds \frac{2\omega}{\sqrt{2\pi}}}\, w_{p,r}(p)_{\kappa} \delta({\bf p}-{\bf q})\end{array}\]since $p$ is on the positive mass shell. These $\ell(x)_\kappa \in {\cal B}$ and the dependence on ${\bf q}$ and $r$ is suppressed in this notation. The polarization of the state created by $b_r^*(q)$ is $w_{p,r}(q)$ from (\ref{pol-st-fermion}). A similar development but for $\Psi(x)$ results in the electron annihilation operator.\[2\omega i \sum_r u_r(q)_{\kappa} b_r(q) =\int \frac{d{\bf x}}{(2\pi)^{3/2}}\; e^{-iqx}(iq_{(0)}\Psi(x)_\kappa+\dot{\Psi}(x)_\kappa).\]The orthogonality relation (\ref{pol-basis}) in this case results in\[b_r(q) =\sum_\kappa \int dp\; \tilde{\Psi}(-p)_\kappa\tilde{\ell}^a(0;p)_\kappa\]for a plane wave limit of LSZ functions\[\tilde{\ell}^a(0;p)_\kappa ={\ds \frac{2\omega}{\sqrt{2\pi}}}\, \overline{w_{p,r}}(p)_{\kappa} \delta({\bf p}-{\bf q}).\]

A general polarization is represented in (\ref{polarzn}) using\begin{equation}\label{pol-fermion-genrl} w_p(p)={\ds \sum_{r=1}^2} c_r\, w_{p,r}(p),\qquad w_a(p)={\ds \sum_{r=1}^2} c_{r+2}\, w_{a,r}(p).\end{equation}When $\sum_{k=1}^4 |c_k|^2=1$, polarization descriptions (\ref{pol-fermion-genrl}) satisfy the normalization (\ref{pold}). From the orthogonality relations (\ref{pol-basis}) and the representation of $DM(p)$ from (\ref{2pt-PP*}) and (\ref{2pt-P*P}),\[\renewcommand{\arraystretch}{1.25} \begin{array}{rl} 2\omega\; \overline{w}(p)^T DM(p) w(p)&=(2\omega)^2 {\ds \sum_{r=1}^2}\, \overline{w}(p)^T \left(\begin{array}{cc} \overline{v_r}(p)v_r(p)^T &0\\ 0 & u_r(p)\overline{u_r}(p)^T\end{array}\right) w(p)\\
 &={\ds \sum_{k=1}^4} |c_k|^2\\
 &=1.\end{array}\]
 
% = = = = = = = = = = = = = = = = = =
\subsection{The Lorentz transformation of $M_2(p)$} \label{app-eqrep}

The Lorentz transformations of $M_2(p)$ are developed using the SL(2,C) representation of the covering group of the Lorentz group. The particular two-point function (\ref{M2-defn}) is selected for contrasts with Feynman perturbation analysis [\ref{weinberg},\ref{schwabl}].

% = = = = = = = = = = = = = = = = = = = =
% *** Lorentz group conventions
% = = = = = = = = = = = = = = = = = = = =
The convention followed in this note is set by (\ref{matcond2}),\[S(A)M(p)S(A)^T = M(\Lambda^{-1} p).\]The representation of the covering group of the Lorentz group is that\[AP(p)A^* = P(\Lambda p)\]for the matrix $P(p)$ defined in (\ref{P-fermion}) and $A\in$SL(2,C) with $\Lambda(A)_{jk}=\frac{1}{2}\mbox{Trace}(\sigma_j A \sigma_k A^*)$.

Only the fermion component $M_2(p)$ from (\ref{matcond}) is considered in this appendix. From (\ref{M2-defn}),\[M_2(p)= 2\pi\, \left({\renewcommand{\arraystretch}{1} \begin{array}{cc}
0 & ({\not \! p}+m)\gamma_0\\
\gamma_0({\not \! p}-m)^T & 0 \end{array}}\right), \qquad D_2=\left( \renewcommand{\arraystretch}{1} \begin{array}{cc} 0 & 1\\
1 & 0\end{array} \right).\](\ref{M2-defn}) satisfies conditions (\ref{matcond}) and (\ref{matlocal}). (\ref{matcond}) is satisfied since in the rest frame, $p=(m,0,0,0)$, $D_2M_2(0)$ is evidently positive semidefinite from (\ref{DM(0)}) and satisfaction of Lorentz covariance provides that $D_2M_2(p)$ is positive semidefinite. Lorentz covariance is a definiteness preserving $*$-congruence of $D_2M_2(0)$. Indeed, $D_2M_2(p)$ has two fourfold degenerate eigenvalues, $\lambda_\pm = E\pm \omega$, and $\lambda_+=2\omega >0$ and $\lambda_-=0$ when $E=\omega$. $\overline{D_2}=D_2$ and $D_2^2=1$, the 8x8 identity. (\ref{matlocal}) is satisfied by (\ref{M2-defn}) with the sign for fermions, $M_2(-p)^T=-M_2(p)$ from ${\not \! p}\mapsto -{\not \! p}$ for $p\mapsto -p$, linearity, the reverse-order law of matrix transposition. $(M_1M_2)^T=M_2^TM_1^T$ and $\gamma_0^T=\gamma_0$.

Development of the Lorentz transformations of $M_2(p)$ uses a similarity transformation of the gamma matrices (\ref{gamma}) to an equivalent representation with an evident implementation of the Lorentz group. Conditions (\ref{matcond}), (\ref{matlocal}), and (\ref{matcond2}) are preserved under simultaneous real, orthogonal similarity transforms of $M_2(p),D_2,S_2(A)$. With\begin{equation}\label{trangam} B=\frac{1}{\sqrt{2}}\left(\begin{array}{cc} \sigma_0 & -\sigma_0\\ \sigma_0 & \sigma_0\end{array}\right), \qquad B^{-1}=\frac{1}{\sqrt{2}}\left(\begin{array}{cc} \sigma_0 & \sigma_0\\ -\sigma_0 & \sigma_0\end{array}\right),\end{equation}and\[\gamma'_\mu =B\gamma_\mu B^{-1}\]from (\ref{gamma}), the equivalent representation of the gamma matrices is\[ \gamma_0'=\left(\begin{array}{cc} 0 & \sigma_0\\\sigma_0 & 0\end{array}\right), \qquad \qquad \gamma_j'=\left(\begin{array}{cc} 0 & \sigma_j\\ -\sigma_j & 0\end{array}\right)\]for $j=1,2,3$. $B$ is a real orthogonal transform. In this representation,\[{\not \! p}\gamma_0'=\left( \renewcommand{\arraystretch}{1} \begin{array}{cc} P(E,-{\bf p}) & 0\\
0 & P(p)\end{array} \right) ,\qquad m\gamma_0'= \left( \renewcommand{\arraystretch}{1} \begin{array}{cc} 0 & m\\
m & 0\end{array} \right)\]from (\ref{slash}) and $P(p)$ from (\ref{P-fermion}). From $AP(p)A^*=P(\Lambda(A) p)$ with $A\in$SL(2,C), the Lorentz transformations are\[\left( \renewcommand{\arraystretch}{1} \begin{array}{cc} (A^*)^{-1} & 0\\
0 & A\end{array} \right)\left( \renewcommand{\arraystretch}{1} \begin{array}{cc} P(E,-{\bf p}) & 0\\
0 & P(p)\end{array} \right)\left( \renewcommand{\arraystretch}{1} \begin{array}{cc} A^{-1} & 0\\
0 & A^*\end{array} \right)=\left( \renewcommand{\arraystretch}{1} \begin{array}{cc} P(E',-{\bf p}') & 0\\
 0 & P(\Lambda(A) p)\end{array} \right).\]$p\,'=\Lambda(A) p$. The transformation of $P(E,-{\bf p})$ follows from\[\renewcommand{\arraystretch}{1.25} \begin{array}{rl}P(E',-{\bf p}') &= \sigma_2 P(p\,')^T \sigma_2\\
 &= \sigma_2 (AP(p)A^*)^T \sigma_2\\
 &= \sigma_2 \overline{A} \sigma_2 \sigma_2 P(p)^T \sigma_2 \sigma_2 A^T \sigma_2\\
 &=(A^*)^{-1} P(E,-{\bf p}) A^{-1},\end{array}\]from $\sigma_2^2=1$, $\sigma_2 A^T \sigma_2= A^{-1}$ and $\sigma_2 \overline{A} \sigma_2= (A^*)^{-1}$ for $A\in$SL(2,C). From (\ref{matcond2}), (\ref{M2-defn}) and (\ref{lorentz-fermion-full}),\[\overline{S_p}(A)^{-1} ({\not \! p}+m)\gamma_0'\,(S_p(A)^T)^{-1}=({\not \! p}\,'+m)\gamma_0'\]with\[\overline{S_p}(A)^{-1}=\left( \renewcommand{\arraystretch}{1} \begin{array}{cc} (A^*)^{-1} & 0\\
0 & A\end{array} \right).\]Transformation of the term proportional to mass results in the useful identity\begin{equation}\label{sp-iden2}\overline{S_p}(A)^{-1} \,\gamma_0'(S_p(A)^T)^{-1}=\gamma_0'\end{equation}valid for real orthogonal similar representations of the gamma matrices.

Returning to the original, unprimed representation of the gamma matrices using (\ref{trangam}),\begin{equation}\label{sp-eval}\overline{S_p}(A) = B^{-1}\left( \renewcommand{\arraystretch}{1} \begin{array}{cc} A^* & 0\\
0 & A^{-1}\end{array} \right)B=\frac{1}{2}\left( \renewcommand{\arraystretch}{1} \begin{array}{cc} A^{-1}+A^* & A^{-1}-A^*\\
A^{-1}-A^* & A^{-1}+A^*\end{array} \right)\end{equation}with the inverse transformation\[\overline{S_p}(A)^{-1}= \overline{S_p}(A^{-1})=\gamma_0 S_p(A)^T \gamma_0 ={\ds \frac{1}{2}} \left( \renewcommand{\arraystretch}{1.25} \begin{array}{cc} A\!+\!(A^*)^{-1} & A\!-\!(A^*)^{-1}\\
A\!-\!(A^*)^{-1} & A\!+\!(A^*)^{-1}\end{array} \right)\]using (\ref{gamma}), (\ref{sp-iden2})  and $(A^{-1})^*=(A^*)^{-1}$. To complete the Lorentz transformation of the fermion two-point function, $\overline{S}(A)DM_2(p)S(A)^T=DM_2(\Lambda^{-1}p)$, use the result above but with a negative mass. With $p\,'=\Lambda(A)p$,\[\renewcommand{\arraystretch}{1.25} \begin{array}{rl} \gamma_0({\not \! p}\,'-m)^T &= (({\not \! p}\,'-m)\gamma_0)^T\\
 &=(\overline{S_p}(A)^{-1} ({\not \! p}-m)\gamma_0 (S_p(A)^T)^{-1})^T\\
 &=S_p(A)^{-1} (({\not \! p}-m)\gamma_0 )^T (S_p(A)^*)^{-1}.\end{array}\]Collecting results, identify\[D_2M_2(\Lambda^{-1}p)= 2\pi\, \left({\renewcommand{\arraystretch}{1} \begin{array}{cc}
S_p(A) & 0\\
 0 & \overline{S_p}(A) \end{array}}\right)
\left({\renewcommand{\arraystretch}{1} \begin{array}{cc}
\gamma_0({\not \! p}\!-\!m)^T & 0\\
 0 & ({\not \! p}\!+\!m)\gamma_0\end{array}}\right)
 \left({\renewcommand{\arraystretch}{1} \begin{array}{cc}
S_p(A)^* & 0\\
 0 & S_p(A)^T\end{array}}\right)\]and (\ref{lorentz-fermion-full}) with (\ref{lorentz-fermion}) are the Lorentz transforms $S_2(A)$. From (\ref{M2-defn}), $\overline{S}_2(A)D_2=D_2S_2(A)$ and (\ref{lorentz-fermion-full}) satisfies (\ref{matcond2}). Finally, $M_2(p),D_2,S_2(A)$ defined in (\ref{M2-defn}) and (\ref{lorentz-fermion-full}) satisfy the conditions (\ref{matcond}), (\ref{matlocal}) and (\ref{matcond2}) and are appropriate for a construction of VEV for spin one-half, charged fermions.

The independence of the components of $p$ and $m$, (\ref{slash}), (\ref{sp-iden2}) and equivalence under real, orthogonal similarity transform provide the identities\begin{equation}\label{sp-iden}\renewcommand{\arraystretch}{1.75} \begin{array}{rl} {\not \! p} &=\overline{S_p}(A) \,{\not \! p}\,'\, \overline{S_p}(A)^{-1}\\
 {\ds \sum_\nu} \gamma_\nu^* \Lambda_{\nu\mu}^{-1} &=\overline{S_p}(A)\gamma_\mu^*\,\overline{S_p}(A)^{-1}\\
 {\ds \sum_\nu} \Lambda_{\mu\nu}^{-1} \gamma_\nu^* &= S_p(A)^T \gamma_\mu^*\,(S_p(A)^T)^{-1}\end{array}\end{equation}with $p=\Lambda(A)^{-1}p\,'$. In the final line, $\Lambda\rightarrow \Lambda^T$ corresponds with $A\rightarrow A^*$ from $\Lambda_{jk}=\frac{1}{2}\mbox{Trace}(\sigma_j A \sigma_k A^*)$ and cyclic invariance of the trace. Then $\overline{S_p}(A^*)=S_p(A)^T$ from (\ref{sp-eval}) results in the identity.

%= = = = = = = = = = = = = = = = = = = = = = =
\subsection{Transition likelihoods and cross sections} \label{app-trans}

The differences of the QFT constructions from a Feynman rules development include that fields are not Hermitian Hilbert space operators and perturbation analyses typically use a box normalization for plane waves. This appendix provides the correspondence of the plane wave limits of the constructed cross sections with cross sections from the Feynman rules perturbation analysis. Plane wave limits of states are not within the Hilbert space, but transition rates are defined in the plane wave limit. This well-known result is briefly reproduced here. 

The likelihood of transition to a scattered state is the trace of the initial state density matrix projected onto the subspace spanned by the final states of interest [\ref{vonN}]. For a system prepared in the pure state $|(p,w)_2^{\mathit{in}}\rangle$, projection onto the subspace of final states near $|(p,w)_{3,4}^{\mathit{out}}\rangle$ gives the likelihood\[\mbox{Trace}(P\rho)= \int \mu(d{\bf p}_3) \mu(d{\bf p}_4)\; \frac{\left| \langle (p,w)_{3,4}^{\mathit{out}} | (p,w)_2^{\mathit{in}} \rangle \right|^2}{\| (p,w)_{3,4}^{\mathit{out}} \|^2 \; \|(p,w)_2^{\mathit{in}} \|^2 }\]with the projection onto final states\begin{equation}\label{proj}P=\int \mu(d{\bf p}_3) \mu(d{\bf p}_4)\; \frac{| (p,w)_{3,4}^{\mathit{out}} \rangle\, \langle (p,w)_{3,4}^{\mathit{out}} |}{\| (p,w)_{3,4}^{\mathit{out}} \|^2 }\end{equation}a summation over momenta in the neighborhood of the ${\bf p}_j$ using a measure determined so that $P^2=P$. The intial state density matrix\[\rho= \frac{| (p,w)_2^{\mathit{in}} \rangle\, \langle (p,w)_2^{\mathit{in}} |}{\|(p,w)_2^{\mathit{in}} \|^2 }\]has Trace$(\rho)=1$. The $|(p,w)_2^{\mathit{in}}\rangle$ are plane wave limits of states with momenta ${\bf p}_1,{\bf p}_2$ and polarizations described by $w_1,w_2$.
%= = = = = = = 

The free field contribution to the four-point function is necessary to evaluation of state norms and is the result of\[\tilde{W}_{o;4}((\xi)_4)= \langle \Omega | \tilde{\Phi}_1 \tilde{\Phi}_2 \tilde{\Phi}_3 \tilde{\Phi}_4 \Omega\rangle \]with $\tilde{\Phi}_i:=a_i+a^*_i+c_i+c^*_i$ the Fourier transforms of free fields composed of creation and annihilation operators satisfying CCR and CAR as $[a_i,a_j^*] :={\cal B}_{ij}$ for bosons, $[c_i,c_j^*]_+ :={\cal F}_{ij}$ for fermions. All other pairings of $a_i,a^*_i,c_i,c^*_i$ commute or anticommute in the case of paired fermionic operators, and $|a_i\,\Omega\rangle = |c_i\,\Omega \rangle =0$. From (\ref{twopoint}) and in ${\cal B}$,\[\tilde{W}_2(\xi_i,\xi_j)={\cal B}_{ij}+{\cal F}_{ij}\]with\[ {\cal B}_{ij} = \delta(p_i+p_j)\delta_j^+ \left( \renewcommand{\arraystretch}{1.25} \begin{array}{cc} M_1(p_j) & 0\\ 0 & 0\end{array} \right) \qquad \qquad {\cal F}_{ij} = \delta(p_i+p_j)\delta_j^+ \left( \renewcommand{\arraystretch}{1.25} \begin{array}{cc} 0 & 0\\ 0 & M_2(p_j)\end{array} \right).\]The four-point functions are the sum of this free field four-point function and the four-point connected function (\ref{trun-eval}).\begin{equation}\label{link4}\renewcommand{\arraystretch}{1.25} \begin{array}{l} \tilde{W}_4((\xi)_4) = {^C\tilde{W}_4((\xi)_4)} + ({\cal B}_{12}+{\cal F}_{12})({\cal B}_{34}+{\cal F}_{34})+({\cal B}_{14}+{\cal F}_{14})({\cal B}_{23}+{\cal F}_{23})\\
 \qquad \qquad \qquad +{\cal B}_{13}{\cal B}_{24}+{\cal B}_{13}{\cal F}_{24}+{\cal F}_{13}{\cal B}_{24}-{\cal F}_{13}{\cal F}_{24}.\end{array}\end{equation}
%= = = = = = = = = = = = = = = = = = = = = = =

The plane wave limit for the two-point VEV derives from the two-point function (\ref{twopoint}) evaluated for LSZ functions (\ref{LSZ}),\begin{equation}\label{2pt-norm}\renewcommand{\arraystretch}{1.75} \begin{array}{l} \langle (p_i,w_i)^{\mathit{in}} | (p_k,w_k)^{\mathit{in}}\rangle ={\ds \int} dq_i dq_k\; \overline{w}_i^T DM(q_k) w_k\; \delta(q_i+q_k) \delta_k^+ \;\left({\ds \frac{L}{\sqrt{\pi}}}\right)^{2(d-1)} \times\\
 \qquad \qquad(\omega_i\!-\!E_i)e^{-i(\omega_i+E_i)t_i} (\omega_k\!+\!E_k)e^{i(\omega_k-E_k)t_k} e^{-L^2({\bf q}_i+{\bf p}_i)^2}e^{-L^2({\bf q}_k-{\bf p}_k)^2}\\
 \qquad = \left({\ds \frac{L}{\sqrt{\pi}}}\right)^{2(d-1)} {\ds \int}d{\bf q}_k \;\overline{w}_i^T DM(q_k) w_k\; 2\omega_k e^{-L^2({\bf q}_k-{\bf p}_i)^2}e^{-L^2({\bf q}_k-{\bf p}_k)^2}\\
 \qquad \approx 2\omega_k \;(\overline{w}_i^T DM(p_k) w_k)\left({\ds \frac{L}{\sqrt{2\pi}}}\right)^{d-1}e^{-L^2({\bf p}_i-{\bf p}_k)^2/2}. \end{array}\end{equation}The approximations follow from the mean value theorem for integration and the plane wave limit. With the normalization (\ref{pold}), $2\omega_k \;(\overline{w}_k^T DM(p_k) w_k)=1$ for each $k$,\[\langle (p_i,w_k)^{\mathit{in}} | (p_k,w_k)^{\mathit{in}}\rangle \approx \delta({\bf p}_i-{\bf p}_k).\]This is the only contributor to the two-point function.\begin{equation}\label{2pt-norm2} \langle \, \underline{1} | (p_i,w_i,p_k,w_k)^{\mathit{in}}\rangle =\langle (p_i,w_i,p_k,w_k)^{\mathit{out}} |\, \underline{1}\rangle=0\end{equation}

(\ref{link4}), (\ref{2pt-norm}) and (\ref{2pt-norm2}) provide the result for norming the states. The contribution of the connected four-point function to state norms is negligible in the plane wave limit as the free field, forward terms dominate, order $L^{2(d-1)}$ to order $L^{d-1}$ from the connected four-point function. When ${\bf p}_i\neq {\bf p}_j$, the argument transposition in (\ref{dualf}) results in\begin{equation}\label{snorm} \renewcommand{\arraystretch}{1.25} \begin{array}{rl} \| (p,w)_{i,j}^{\mathit{in}} \|^2 &\approx \langle (p_i,w_i)^{\mathit{in}} | (p_i,w_i)^{\mathit{in}}\rangle \langle (p_j,w_j)^{\mathit{in}} | (p_j,w_j)^{\mathit{in}}\rangle\\
 &= 4\omega_i\omega_j \;(\overline{w}_i^T DM(p_i) w_i) \;(\overline{w}_j^T DM(p_j) w_j)\left({\ds \frac{L^2}{2\pi}}\right)^{d-1}\\
 &= \left({\ds \frac{L^2}{2\pi}}\right)^{d-1}\end{array}\end{equation}from the normalization (\ref{pold}). Results for $\mathit{in}$ also apply for $\mathit{out}$.

The measure on state labels is evaluated using $P^2=P$. In the plane wave limit, (\ref{proj}), (\ref{2pt-norm}), (\ref{snorm}), and neglecting ${\bf p}_3 = {\bf p}_4$ as measure zero, idempotence of $P$ results in\[ \renewcommand{\arraystretch}{2.25}\begin{array}{rl} P^2
 &= {\ds \int} \mu(d{\bf p}_3) \mu(d{\bf p}_4) {\ds \int} \mu(d{\bf p'}_3) \mu(d{\bf p'}_4) 
{\ds \frac{\langle (p,w)_{3,4}^{\mathit{out}} |(p',w)_{3,4}^{\mathit{out}} \rangle}
{\| (p,w)_{3,4}^{\mathit{out}} \|^2 \| (p',w)_{3,4}^{\mathit{out}} \|^2 }}\,| (p,w)_{3,4}^{\mathit{out}} \rangle\,\langle (p',w)_{3,4}^{\mathit{out}} | \\
 &\approx {\ds \int} \mu(d{\bf p}_3) \mu(d{\bf p}_4) \mu(d{\bf p'}_3) \mu(d{\bf p'}_4) \left({\ds \frac{2\pi}{L^2}}\right)^{d-1} \delta({\bf p}_3\!-\!{\bf p'}_3) \delta({\bf p}_4\!-\!{\bf p'}_4) {\ds \frac{| (p,w)_{3,4}^{\mathit{out}} \rangle\,\langle (p,w)_{3,4}^{\mathit{out}} |}
{\| (p,w)_{3,4}^{\mathit{out}} \|^2 }} \end{array}\]and the identification\[\mu(d{\bf p}_k) = \left( \frac{L}{\sqrt{2\pi}}\right)^{d-1} \; d{\bf p}_k.\]

The differential cross section for two-in, two-out scattering is defined by the likelihood of the plane wave $\mathit{in}$ state scattering into an $\mathit{out}$ state within a momentum increment $d{\bf p}_3 d{\bf p}_4$,\begin{equation}\label{cross} d\sigma = A\,\mbox{Trace}(P\rho)\end{equation} with a flux corrected interaction area\[A := \frac{\ds V}{\ds T\;u_\alpha}\]with $V$ the interaction volume, $T$ the duration of the wave packets, and a velocity [\ref{weinberg}]\[u_\alpha := \frac{\ds \sqrt{(p_1 p_2)^2-m_1^2m_2^2}}{\ds \omega_1\omega_2}.\]

These results, collected together in (\ref{cross}) in the normalization (\ref{pold}), result in the differential cross section for non-forward two-in, two-out scattering into a momenta increment $d{\bf p}_3 d{\bf p}_4$.\[\renewcommand{\arraystretch}{2.25} \begin{array}{rl} d\sigma &= d{\bf p}_3 d{\bf p}_4 \;A\; \left( \frac{\ds L^2}{\ds 2 \pi}\right)^{d\!-\!1}\; \frac{\ds \delta_T(p_1+p_2-p_3-p_4;L^2/4)^2 \;(2\pi)^2|{\cal M}((p,w)_4)|^2}{\ds \| (p,w)_2^{\mathit{in}} \|^2 \; \|(p,w)_{3,4}^{\mathit{out}} \|^2 }\\ 
 &= d{\bf p}_3 d{\bf p}_4 \; {\ds \frac{(2\pi)^d}{u_\alpha}}\,{\ds \frac{V}{(2L\sqrt{\pi})^{d-1}}}\,\delta_T(p_1+p_2-p_3-p_4;L^2/4) \; |{\cal M}((p,w)_4)|^2. \end{array} \]From (\ref{delta}),\[\left( \delta_T(p;L^2/4) \right)^2= \left( \frac{\ds T}{\ds 2\pi} \right)\left( \frac{\ds L}{\ds 2\sqrt{\pi}} \right)^{d-1} \delta_T(p;L^2/4).\]The volume in the flux calculation is\[V=(2L\sqrt{\pi})^{d-1},\]determined consistently with box normalization using\[\delta({\bf p})^2 = \frac{V}{(2\pi)^{d-1}}\,\delta({\bf p})= \left(\frac{L}{\sqrt{\pi}}\right)^{d-1}\delta({\bf p};L^2)\]for the delta sequences (\ref{LSZdelta}).

Integration over all ${\bf p}_4$ and all magnitudes for ${\bf p}_3$ results in the cross section for two particle to two particle scattering into a cone $d\Omega$ without regard to the energy of the scattered product nor the energy and direction of the second product particle. With $\varrho_j^2:={\bf p}_j^2$ and in the center of momentum frame, ${\bf p}_1+{\bf p}_2=0$, $u_\alpha=\varrho_1(\omega_3+\omega_4)/(\omega_1\omega_2)$ using conservation of energy,\[\delta(\omega_1+\omega_2-\omega_3-\omega_4)=\frac{\omega_3\omega_4}{\varrho_3\,(\omega_3+\omega_4)}\; \delta(\varrho_3-\varrho_o)\]with\begin{equation}\label{varrho}\varrho_o:=\frac{\sqrt{((\omega_1+\omega_2)^2-m_1^2-m_2^2)^2-4m_1^2m_2^2}}{2(\omega_1+\omega_2)}.\end{equation}Then in the center of momentum frame,\begin{equation}\label{elascs}\renewcommand{\arraystretch}{2.25} \begin{array}{rl} \frac{\ds d\sigma}{\ds d\Omega} =& {\ds \int_0^{\infty}} \varrho_3^{d-2}d\varrho_3 d{\bf p}_4\; A\mbox{Trace}(P\rho)\\ 
 =& \frac{\ds (2\pi)^d \varrho_o^{d-3} \, \omega_1\omega_2\omega_3\omega_4|{\cal M}((p,w)_4)|^2}{\ds \varrho_1\, (\omega_3+\omega_4)^2}\end{array}\end{equation}evaluated at $\varrho_3=\varrho_o$.
 
%= = = = = = = = = = = = = = = = = = = = = = =
%= = = = = = = = References= = = = = = = = = =
%= = = = = = = = = = = = = = = = = = = = = = =

\section*{References}
\begin{enumerate}
\item \label{gej05} G.E.~Johnson, ``Algebras without Involution and Quantum Field Theories'', 13 March, 2012, arXiv:math-ph/1203.2705v1.
\item \label{mp01} G.E.~Johnson, ``Massless Particles in QFT from Algebras without Involution'', 22 May, 2012, arXiv:math-ph/1205.4323v1.
\item \label{pct} R.F.~Streater and A.S.~Wightman, {\em PCT, Spin and Statistics, and All That}, Reading, MA: W.A.~Benjamin, 1964.
\item \label{bogo} N.N.~Bogolubov, A.A.~Logunov, and I.T.~Todorov, {\em Introduction to Axiomatic Quantum Field Theory}, trans.~by Stephen Fulling and Ludmilla Popova, Reading, MA: W.A.~Benjamin, 1975.
\item \label{wightman-hilbert} A.S.~Wightman,``Hilbert's Sixth Problem: Mathematical Treatment of the Axioms of Physics'', {\em Mathematical Development Arising from Hilbert Problems}, ed.~by F.~E.~Browder, {\em Symposia in Pure Mathematics 28}, Providence, RI: Amer.~Math.~Soc., 1976, p.~147.
\item \label{wight} A.S.~Wightman, ``Quantum Field Theory in Terms of Vacuum Expectation Values'', {\em Phys.~Rev.}, vol.~101, 1956, p.~860.
\item \label{borchers} H.J.~Borchers, ``On the structure of the algebra of field operators'', {\em Nuovo Cimento}, Vol.~24, 1962, p.~214.
\item \label{wigner} T.D.~Newton and E.P.~Wigner, ``Localized States for Elementary Systems'', {\em Rev. Modern Phys.}, Vol.~21, 1949, p.~400.
%\item \label{baum} H.~Baumg\"{a}rtel, and M.~Wollenberg, ``A Class of Nontrivial Weakly Local Massive Wightman Fields with Interpolating Properties'', {\em Commun. Math. Phys.}, Vol.~94, 1984, p.~331.
%\item \label{longo} R.~Brunette, D.~Guido, and R.~Longo, ``Modular Localization and Wigner Particles'', {\em Commun. Math. Phys.}, arXiv:math-ph/0203021v2.
%\item \label{yngvason} J. Yngvason, ``The Borchers-Uhlmann Algebra and its Descendants'', G\"{o}ttingen, July 29, 2009, www.lqp.uni-goettingen.de/events/aqft50/slides/1-4-Yngvason.pdf.
\item \label{yngvason} J. Yngvason, ``On the algebra of test functions for field operators'', {\em Commun. Math. Phys.}, Vol. 34, 1973, p. 315.
\item \label{segal} I.E.~Segal and R.W,~Goodman, ``Anti-locality of certain Lorentz-invariant operators'', {\em Journal of Mathematics and Mechanics}, Vol.~14, 1965, p.~629.
\item \label{gel2} I.M.~Gel'fand, and G.E.~Shilov, {\em Generalized Functions}, Vol.~2, trans.~M.D.~Friedman, A.~Feinstein, and C.P.~Peltzer, New York, NY: Academic Press, 1968.
\item \label{feder} P.G.~Federbush and K.A.~Johnson, ``The Uniqueness of the Two-Point Function'', {\em Phys. Rev.}, Vol.~120, 1960, p.~1926.
\item \label{greenberg} O.W.~Greenberg, ``Heisenberg Fields which vanish on Domains of Momentum Space'', {\em Journal of Math.~Phys.}, Vol.~3, 1962, pp.~859-866.
%\item \label{strocchi2} F.~Strocchi, ``Extension of Jost and Schroer's Theorem to Quantum Electrodynamics'', {\em Phys. Rev. D}, Vol. 6, Aug. 1972, p. 1193.
%\item \label{mund} J.~Mund, ``An Algebraic Jost-Schroer Theorem for Massive Theories'', arXiv:hep-ph/1012.1454v3, Dec. 2010.
%\item \label{heger} G.C. Hegerfeldt, ``Prime Field Decompositions and Infinitely Divisible States on Borchers' Tensor Algebra'', {\em Commun. Math. Phys.}, Vol.~45, 1975, p.~137.
%\item \label{rinke} H.~Rinke, ``A remark on the asymptotic completeness of local fields'', {\em Commun. Math. Phys.}, Vol.~12, 1969, p.~324.
\item \label{steinmann} O.~Steinmann, ``Structure of the Two-Point Function'', {\em Journal of Math. Phys.}, Vol. 4, 1963, p. 583.
\item \label{ewg} B.S.~DeWitt, H.~Everett III, N.~Graham, J.A.~Wheeler republished in {\em The Many-worlds Interpretation of Quantum Mechanics}, ed.~B.S.~DeWitt, N. Graham, Princeton, NJ: Princeton University Press, 1973.
\item \label{vonN} J.~von Neumann, {\em Mathematical Foundations of Quantum Mechanics}, Princeton, NJ: Princeton University Press, 1955.
\item \label{gleason} A.M.~Gleason, ``Measures on the closed subspaces of a Hilbert space'', {\em Journal of Mathematics and Mechanics}, Vol.~6, 1957, p.~885.
\item \label{birk} G.~Birkhoff and J.~von Neumann, ``The Logic of Quantum Mechanics'', {\em Ann. Math.}, Vol.~37, 1936, p. 823.
\item \label{dirac} P.A.M.~Dirac, {\em The Principles of Quantum Mechanics, Fourth Edition}, Oxford: Clarendon Press, 1958.
\item \label{gel4} I.M.~Gel'fand, and N.Ya.~Vilenkin, {\em Generalized Functions}, Vol.~4, trans.~A.~Feinstein, New York, NY: Academic Press, 1964.
\item \label{type3} J.~Yngvason, ``The Role of Type III Factors on Quantum Field Theory'', arXiv:math-ph/0411058v2 [math-ph], Dec. 2004.
%\item \label{emch} G.E.~Emch, {\em Algebraic Methods in Statistical Mechanics and Quantum Field Theory}, New York, NY: Wiley-Interscience, 1972.
\item \label{svn} J.~Rosenberg, ``A selective history of the Stone-von Neumann theorem'', in {\em Operator Algebras, Quantization, and Noncommutative Geometry: A Centennial Celebration Honoring John von Neumann and Marshall H. Stone}, R.S.~Doran and R.V.~Kadison ed., American Mathematical Society, 2004.
\item \label{messiah} A.~Messiah, {\em Quantum Mechanics}, vol.~1, New York, NY: John Wiley and Sons, 1968.
\item \label{jackson} J.D.~Jackson, {\em Classical Electrodynamics}, New York, NY: John Wiley and Sons, 1962.
\item \label{weinberg} S.~Weinberg, {\em The Quantum Theory of Fields, Volume I, Foundations}, Cambridge: Cambridge University Press, 2005.
%\item \label{gross} F.~Gross, {\em Relativistic Quantum Mechanics and Field Theory}, New York, NY: Wiley-Interscience Publication, 1993.
\item \label{merzbacher} Eugen Merzbacher, {\em Quantum Mechanics, Second Edition}, New York, NY: John Wiley and Sons, 1970.
\item \label{rgnewton} Roger~G.~Newton, {\em Scattering Theory of Waves and Particles}, New York, NY: McGraw-Hill, 1966.
\item \label{schwabl} F.~Schwabl, {\em Advanced Quantum Mechanics}, trans.~R.~Hilton and A.~Lahee, Berlin: Springer-Verlag, 1999.
\item \label{horn} R.A. Horn, and C.R. Johnson, {\em Matrix Analysis}, Cambridge: Cambridge University Press, 1985.
\item \label{combin} M. Hardy, ``Combinatorics of Partial Derivatives'', {\em The Electronic Journal of Combinatorics}, Vol. 13, \#R1, 2006.
\item \label{nbs} {\em Handbook of Mathematical Functions}, ed. M. Abramowitz, and I.A. Stegun, National Bureau of Standards, Applied Mathematics Series - 55, 1970.
\item \label{cook} J.M.~Cook, ``The Mathematics of Second Quantization'', {\em Trans.~Am.~Math.~Soc.}, Vol.~74, 1953, pg.~222.
% = = = 
\item \label{haagasym} R.~Haag, ``Quantum Field Theories with Composite Particles and Asymptotic Conditions'', {\em Phys.~Rev.}, Vol.~112, 1958, p.~669.
\item \label{ahh} H.~Araki, K.~Hepp, and R.~Haag, ``On the Asymptotic Behavior of Wightman Functions in Space-Like Directions'', {\em Helv.~Phys.~Acta.}, Vol.~35, 1962, p.~164.
\end{enumerate}
\end{document}